\documentclass[12pt]{amsart}
\usepackage{graphicx,amsmath,amssymb,epsfig}
\usepackage{amsfonts, lscape, harvard, chicago}
\usepackage{fancyhdr}
\usepackage{lipsum}

\usepackage{hyperref}

\hypersetup{
    colorlinks=true,
    urlcolor=blue,
    linkcolor=blue,
    citecolor=blue
}

\newtheorem{theorem}{Theorem}
\theoremstyle{plain}

\newtheorem{example}{Example}

\newtheorem{lemma}{Lemma}

\numberwithin{equation}{section}
\input{tcilatex}

\newcommand\blfootnote[1]{%
  \begingroup
  \renewcommand\thefootnote{}\footnote{#1}%
  \addtocounter{footnote}{-1}%
  \endgroup
}

\begin{document}

\setcounter{page}{1}

\title{Personality traits and the marriage market}

\author{ARNAUD DUPUY\\
        CEPS/INSTEAD, Maastricht School of Management and IZA \\
\and \\
ALFRED GALICHON\\
       Sciences Po, Paris, IZA and CEPR.
}

\maketitle

\begin{abstract}
Which and how many attributes are relevant for the sorting of agents in a matching market? This paper addresses these questions by constructing indices of mutual attractiveness that aggregate information about agents' attributes. The first k indices for agents on each side of the market provide the best approximation of the matching surplus by a k-dimensional model. The methodology is applied on a unique Dutch households survey containing information about education, height, BMI, health, attitude toward risk and personality traits of spouses.
\end{abstract}

\blfootnote{We thank five anonymous referees, the Editor (Phil
Reny), as well as Raicho Bojilov, Odran Bonnet, Xavier Gabaix, Jim Heckman,
Zuzanna Kosowska-Stamirowska, Jean-Marc Robin, Marko Tervi\"{o}, Bertrand
Verheyden, Simon Weber and seminar participants at the University of
Chicago, Universit\'{e} de Montr\'{e}al, Paris 1 Panth\'{e}on--Sorbonne,
University of Alicante, Tilburg University, Sciences Po, Harvard-MIT,
Universit\'{e} de Lausanne, and the 2013 EEA meeting for their comments and
Lex Borghans for useful discussions about the DNB data. Jinxin He provided
excellent research assistance. Dupuy warmly thanks ROA at Maastricht
University where part of this paper was written. Galichon's research has
received funding from the European Research Council under the European
Union's Seventh Framework Programme (FP7/2007-2013) / ERC grant agreement no 313699, and from FiME, Laboratoire de Finance des March\'{e}s de l'Energie.

Accepted for publication by the \textit{Journal of Political Economy}
Volume 122, Number 6, December 2014. Article DOI: \url{https://doi.org/10.1086/677191}.}

\newpage

\section{Introduction}

\setcounter{page}{2}\setcounter{equation}{0}

Marriage, understood in a broad sense, is probably one of the most important
factors for happiness\footnote{%
See e.g. Stutzer and Frey (2006) and Zimmermann and Easterlin (2006).}. It
also plays an important role in the generation of welfare and its
redistribution across individuals. An in-depth understanding of marriage
patterns is therefore of crucial importance for the study of a wide range of
economic issues. A growing body of the economic literature studies the
determinants of marriage, seen as a competitive matching market, both
empirically and theoretically. This literature draws insights from the
seminal model of the marriage market developed by Becker (1973). At the
heart of Becker's theory lies a two-sided assignment model with transferable
utility where agents on both sides of the market (men and women) are
characterized by a set of attributes only partly observed by the researcher.
Each agent aims at matching with a member of the opposite sex so as to
maximize his or her own payoff. This model is particularly interesting since
under certain conditions, one can identify and estimate features of agents'
preferences. A central question in this market is which and how many
attributes are relevant for the sorting of agents?

A large body of literature\footnote{%
For the marriage markets, see among others Becker (1991), Wong (2003),
Anderberg (2004), Choo and Siow (2006), Browning et al. (2013), Chiappori
and Oreffice (2008), Hitsch et al. (2010), Chiappori et al. (2010),
Chiappori et al. (2012), Oreffice and Quintana-Domeque (2010), Bruze (2011),
Charles et al. (2013), Echenique et al. (2013), and Jacquemet and Robin
(2013). For other markets, see e.g. Fox (2010, 2011), Tervi\"{o} (2008),
Gabaix and Landier (2008).} has focused on the identification and estimation
of preferences in the marriage market and in other matching markets;
however, it has been constrained by some methodological limitations
regarding the quantitative methods available to identify and estimate
features of the joint utility function. In the current state of the art, no
estimation tool can handle sorting on multiple continuous attributes in a
convenient manner. Until recently, most empirical literature assumed sorting
occurs on a single continuous dimension, which is a single index aggregating
the various attributes of the agents. The choice of this approach was
strongly influenced by Becker's seminal model of \textquotedblleft positive
assortive mating\textquotedblright , which is essentially
single-dimensional. Due to this limitation, empirical studies to date have
therefore either focused on one attribute at a time, hence ignoring the
effect of other attributes on sorting (see e.g. Charles et al., 2013), or
assumed that all observed attributes matter but only through a single index
of mutual attractiveness (see e.g. Wong, 2003, Anderberg, 2004 and
Chiappori, Oreffice and Quintana-Domeque, 2012). More recently however, a
new vein of the literature initiated by Choo and Siow (2006), and pursued by
Fox (2010, 2011), Chiappori et al. (2013), Galichon and Salani\'{e} (2010,
2013) among others, has built on discrete choice theory and is therefore
restricted to the case of discrete characteristics. It seems fair to assess
that a standard procedure for the estimation of continuous multivariate
matching models is still needed, in spite of recent attention on the matter%
\footnote{%
Recently, two papers have studied markets where sorting occurs on more than
one dimension. Coles and Francesconi (2011) and Chiappori et al. (2012)
study sorting on a single continuous index and a binary variable. Nesheim
(2012) focuses on the identification of multivariate hedonic models without
heterogeneity, and based on the observation of the price.}. Another
limitation of the current empirical literature is related to the set of
observable attributes available in the data. Most studies solely have access
to data on education and earnings, and only a few observe other dimensions
such as anthropometric measures captured by height and BMI or self-assessed
measures of health (Chiappori, Oreffice and Quintana-Domeque, 2012 and
Oreffice and Quintana-Domeque, 2010 are notable exceptions).

In the present paper, we contribute to the literature on three accounts.

\label{justifContinuousChooSiow}First, on the modeling front, we extend: (i)
the Choo and Siow matching model to account for possibly continuous
multivariate attributes, and (ii) Galichon and Salani\'{e}'s (2010, 2013)
surplus estimator of the Choo and Siow model to the continuous case\footnote{%
None of these two papers allows for continuous observable characteristics.}.
Extending Choo and Siow's model to continuous regressors is an important
problem, which has been left open so far. Indeed, many attributes that
appear in empirical studies on the marriage market are intrinsically
continuous: income, wealth, height, body mass index and, as our paper
illustrates, psychometric attributes such as personality traits. Even if
measuring necessarily involves discretization, it remains desirable to have
models which treat attributes as continuous. Using the Choo and Siow model
directly on the discretized attributes to perform inference is problematic
since changing the level of discretization of the data will imply modifying
the assumptions of the model. To solve this problem, we make use of a
continuous version of the logit choice framework, pioneered by Cosslett
(1988) and Dagsvik (1994), which relies on extreme value stochastic
processes. This ensures that our assumptions do not depend on the level of
discretization of the data.

Second, on the data analysis front, we introduce a new technique, which we
call \textquotedblleft Saliency Analysis,\textquotedblright\ to determine
the most relevant dimensions on which sorting occurs in a matching market.
The starting point of this analysis requires inferring the strength of
complementarities between men and women's attributes. Using our structural
model, we evaluate the intensity of assortativeness (positive or negative)
between any pair of attributes, and we call the resulting matrix
\textquotedblleft affinity matrix.\textquotedblright\ Saliency Analysis
consists in analysing the affinity matrix by means of a Singular Value
Decomposition. This allows one to derive \textquotedblleft indices of mutual
attractiveness,\textquotedblright\ such that the joint utility of matching
is a sum of mutually exclusive pairwise interaction terms. The first $k$
indices (for males and females) provide a convenient approximation of the
joint utility by a model where attributes are vectors of only $k$
dimensions. As a consequence, one can perform inference on the number of
dimensions that are required to explain the equilibrium sorting by testing
how many singular values differ from zero.

Third, on the empirical front, we make use of a dataset that allows us to
observe a wide range of attributes of both spouses. The set of attributes we
observe in the data includes socio-economic variables such as education,
anthropometric measures such as height and BMI, a measure of self-assessed
health, as well as psychometric attributes such as risk aversion and the
\textquotedblleft Big Five\textquotedblright\ personality traits well-known
in Psychology: conscientiousness, extraversion, agreeableness, emotional
stability and autonomy. This paper is, to the extent of our knowledge, the
first attempt to evaluate the importance of personality traits in the
sorting of men and women in the marriage market. We will show that although
education explains 28\% of a couple's observable joint utility, personality
traits explain another 17\% and different personality traits matter
differently for men and for women. Our results relate to the literature
showing the importance of personality traits in making economic decisions
(Borghans et al., 2008 for instance). Bowles et al. (2001) and Mueller and
Plug (2006) among others have shown the importance of personality traits for
earnings inequality. Closer to our focus, Lundberg (2012) studies the impact
of personality traits on the odds in and out of a relationship (marriage and
divorce) and finds empirical evidence that personality traits significantly
affect the extensive margin in the marriage market. In particular,
conscientiousness increases the probability of marriage at the age of 35 for
men and extraversion increases the odds of marriage at the age of 35 for
women. In the present work, we study the intensive margin, that is to whom
conscientious men and extraverted women are the most attractive. We show
among other things that conscientious men have preferences for conscientious
women whereas extraverted women have preferences for autonomous and less
agreeable men.

The outline of the rest of the paper is as follows. Section \ref%
{sec:ContinuousChooSiow} presents an important extension of the model of
Choo and Siow to continuously distributed observables. Section \ref%
{sec:param} deals with parametric estimation of the joint utility function
in this setting. Section \ref{sec:Saliency} presents a methodology for
deriving \emph{indices of mutual attractiveness} that determine the
principal dimensions on which sorting occurs. The problem of inferring the
number of dimensions on which sorting occurs is dealt with in section \ref%
{sec:Asympt}. Section \ref{sec:Data} presents the data used for our
empirical estimation and Section \ref{sec:EmpiricalResults} discusses the
results. Section \ref{sec:Conc} concludes.

\newpage

\section{The Continuous Choo and Siow model\label{sec:ContinuousChooSiow}}

\subsection{The Becker-Shapley-Shubik model of marriage}

The setting is a one-to-one, bipartite matching model with transferable
utility. Men and women are characterized by vectors of attributes,
respectively denoted $x\in \mathcal{X}=\mathbb{R}^{d_{x}}$ for men and $y\in 
\mathcal{Y}=\mathbb{R}^{d_{y}}$ for women. Matched men and women are by
definition in equal number; we let $P$ and $Q$ be the respective probability
distributions of their attributes. Throughout the paper, $P$ and $Q$ are
treated as exogenous, except in Appendix~\ref{app:Singles} where we show
that incorporating singles leaves the analysis unchanged while allowing us
to identify reservation utilities. $P$ and $Q$ are assumed to have densities
with respect to the Lebesgue measure\footnote{%
While we present the case with continuous distributions for $x$ and $y$, our
framework easily extends to the case where some dimensions of $x$ and $y$
are discrete.} denoted respectively $f$ and $g$. Without loss of generality,
it is assumed throughout that $P$ and $Q$ are centered distributions, that
is $\mathbb{E}_{P}\left[ X\right] =\mathbb{E}_{Q}\left[ Y\right] =0$.

\bigskip

A \emph{matching} is the probability density $\pi \left( x,y\right) $ of
occurrence of a couple with characteristics $\left( x,y\right) $ from the
matched population. Quite obviously, this imposes that the marginals of $\pi 
$ should be $P$ and $Q$. Write $\pi \in \mathcal{M}\left( P,Q\right) $,
where 
\begin{equation*}
\mathcal{M}\left( P,Q\right) =\left\{ \pi :\pi \left( x,y\right) \geq 0\text{%
, }\int_{\mathcal{Y}}\pi \left( x,y\right) dy=f\left( x\right) \text{ and }%
\int_{\mathcal{X}}\pi \left( x,y\right) dx=g\left( y\right) \right\} .
\end{equation*}

\bigskip

Let $\Phi \left( x,y\right) $ be the joint utility generated when a man $x$
and a woman $y$ match, which is shared endogenously between them. Let $\Phi
\left( x,\emptyset \right) $ and $\Phi \left( \emptyset ,y\right) $ be the
utility of man $x$ and woman $y$ respectively if they remain single; in
Appendix~\ref{app:Singles}, we shall show that $\Phi \left( x,\emptyset
\right) $ and $\Phi \left( \emptyset ,y\right) $ are identified if and only
if the populations of singles are observed, but that the identification of $%
\Phi \left( x,y\right) $ is not impeded if singles are not observed\footnote{%
In Appendix~\ref{app:Singles}, this will be shown to be a consequence of the
Independence of Irrelevant Alternatives (IIA) property of the logit model.}.
In the rest of the paper we shall assume that only the matched population is
observed, so we will not focus on $\Phi \left( x,\emptyset \right) $ and $%
\Phi \left( \emptyset ,y\right) $; \label{page:addedRemark}as a result, the
matching \emph{surplus} $\Phi \left( x,y\right) -\Phi \left( x,\emptyset
\right) -\Phi \left( \emptyset ,y\right) $ will not be identified. Shapley
and Shubik (1972) have shown that the equilibrium matching $\pi $ maximizes
the total utility%
\begin{equation}
\max_{\pi \in \mathcal{M}\left( P,Q\right) }\mathbb{E}_{\pi }\left[ \Phi
\left( X,Y\right) \right] .  \label{optPi2}
\end{equation}%
Optimality condition (\ref{optPi2}) leads to very strong restrictions%
\footnote{%
A basic result in the theory of optimal transportation (Brenier's theorem)
implies that when $\Phi \left( x,y\right) =x^{\prime }Ay$, the optimal
matching is characterized by $\left( AY\right) _{i}=\partial V\left(
X\right) /\partial x_{i}$ where $V$ is some convex function. Hence as soon
as $A$ is invertible, the matching is \emph{pure}, in the sense that no two
men of the same type may marry women of different types. This is obviously
never observed in the data.} on $\left( X,Y\right) $, which are rarely met
in practice. We need to incorporate some amount of unobserved heterogeneity
in the model.

\subsection{Adding heterogeneities\label{par:heterog}}

\label{page:ModelPresentation}Bringing the model to the data requires the
additional step of acknowledging that sorting might also occur on attributes
that are unobserved to the econometrician. In the case where men and women's
attributes are \emph{discrete}, Choo and Siow (2006) introduced unobservable
heterogeneities into the matching problem by considering that if a man $m$
of attributes $x_{m}=x$ and a woman $w$ of attributes $y_{w}=y$ match, they
create a joint utility $\Phi \left( x,y\right) +\varepsilon _{m}\left(
y\right) +\eta _{w}\left( x\right) $, where $\varepsilon _{m}\left( y\right) 
$ and $\eta _{w}\left( x\right) $ are unobserved random \textquotedblleft
sympathy shocks\textquotedblright\ drawn by individuals. Assuming that $%
\left( \varepsilon _{m}\left( y\right) \right) _{y}$ and $\left( \eta
_{w}\left( x\right) \right) _{x}$ have i.i.d. centered Gumbel (extreme value
type I) distributions with scaling parameter $\sigma /2$, Choo and Siow have
shown that the joint utility $\Phi \left( x,y\right) $ can be split into $%
\Phi \left( x,y\right) =U\left( x,y\right) +V\left( x,y\right) $ such that
the utility of man $m$ matching with a woman of type $y$ is given by%
\begin{equation*}
U\left( x,y\right) +\varepsilon _{m}\left( y\right)
\end{equation*}%
with a similar expression for the utility of woman $w$. An important
implication of this setting is that at equilibrium, agents are indifferent
between partners with same observable attributes: the matching utility of
man $m$ at equilibrium depends only on the observable attributes of that
woman. As a consequence, each agent in the market solves a discrete choice
problem.

\label{justifContinuousChooSiow2}In the Choo and Siow model, partners are
assumed to have i.i.d. Gumbel sympathy shocks for the discrete attributes of
the opposite side of the market. However, in many applied settings, these
attributes are continuous random vectors, and even though the data that the
analyst handles are obviously discretized, there is a strong need for a
continuous framework. To illustrate, we shall take a setting where only the
height of the partners is relevant, and assume that the precision of the
measure is poor, say it is rounded at the nearest foot. A direct implication
of the Choo and Siow assumptions is that individuals' sympathy shocks are
perfectly correlated within a foot bracket, and perfectly independent across
feet. Suppose instead that height is measured at the nearest inch, Choo and
Siow's assumptions would now imply that individuals' sympathy shocks are
perfectly correlated within an inch bracket, and perfectly independent
across inches, which of course comes at odds with the previous assumptions.
So, while it is of course always possible to apply the Choo and Siow setting
to the discretized data, this implicitly leads to ad hoc assumptions which
depend on the level of discretization of the available data.

In the present paper, we shall present an application where $x$ and $y$
measure height, BMI and various personality traits, which have a continuous
multivariate distribution. Hence we need to model the random processes for $%
\varepsilon _{m}\left( x\right) $ and $\eta _{w}\left( y\right) $
accordingly. A legitimate candidate in the wake of Choo and Siow's approach
is the continuous logit model. Although very natural and particularly
tractable, this setting has been surprisingly little used in economic
modeling, with some notable exceptions. McFadden (1976) initiated the
literature of continuous logit models by extending the definition of
Independence of Irrelevant Alternatives (IIA)\ beyond finite sets. Ben-Akiva
and Watanatada (1981) and Ben-Akiva, Litinas and Tsunekawa (1985) define
continuous logit models by taking the limits of the discrete choice
probabilities, with applications in particular to the context of spatial
choice models. Cosslett (1988) and Dagsvik (1988) have independently
suggested using max-stable processes to model continuous choice. We base our
approach on their insights.

Assume that each man $m$ of type $x_{m}=x$ only knows a random subset of the
total population of women we will call \textquotedblleft
acquaintances\textquotedblright , and that man $m$ only considers potential
partners from his set of acquaintances. These acquaintances are indexed%
\footnote{%
As explained in Appendix~\ref{app:ContinuousLogit}, it will result from the
distributional assumptions that each man draws an infinite but countable
number of acquaintances almost surely, so that these can be indexed by the
set of integers.} by $k\in \mathbb{N}$; and their observable attributes are
represented by $y_{k}^{m}$. Each of these acquaintances is associated with a
random \textquotedblleft sympathy shock\textquotedblright\ $\varepsilon
_{k}^{m}$ which enters additively into the man's utility, so that the
utility of a man $m$ who marries a woman of attributes $y_{k}^{m}$ can be
written as 
\begin{equation}
U\left( x,y_{k}^{m}\right) +\frac{\sigma }{2}\varepsilon _{k}^{m},
\label{utMM}
\end{equation}%
where $U\left( x,y\right) $ is the \textquotedblleft
systematic\textquotedblright\ (in Choo and Siow's term) part of the utility
obtained by man $x$ matching with a woman with attributes $y$, whose
existence and characterization will be provided in Theorem~\ref%
{thm:continuous-CS} below. Note that in contrast with the original setting
of Choo and Siow described above, men do not have access to the whole
population of women, but only to their randomly selected set of
acquaintances, which is a subset of the whole population.

We have yet to specify the distribution of $y_{k}^{m}$ and $\varepsilon
_{k}^{m}$. Following Cosslett and Dagsvik's idea, we assume that $\{\left(
y_{k}^{m},\varepsilon _{k}^{m}\right) ,k\in \mathbb{N}\}$ are the points of
a Poisson process on $\mathcal{Y}\times \mathbb{R}$ of intensity $dy\times
e^{-\varepsilon }d\varepsilon $. This means that: (i) the probability that
man $m$ has an acquaintance whose observable attributes are in a small set
of infinitesimal size $dy$ around $y$ and with sympathy shock in a set of
infinitesimal size $d\varepsilon $ around $\varepsilon $ is equal to $%
e^{-\varepsilon }d\varepsilon dy$, and (ii) letting $S$ and $S^{\prime }$ be
two disjoint subsets of $\mathcal{Y}\times \mathbb{R}$, the events
\textquotedblleft $m$ has an acquaintance in $S$\textquotedblright\ and
\textquotedblleft $m$ has an acquaintance in $S^{\prime }$%
\textquotedblright\ are independent. According to the standard theory of
Poisson point processes, this implies that, for $S$ a subset of $\mathcal{Y}%
\times \mathbb{R}$, the probability that man $m$ has no acquaintance in set $%
S$ is $\exp \left( -\int_{S}e^{-\varepsilon }dyd\varepsilon \right) $. In
Appendix~\ref{app:ContinuousLogit} we show that this yields a continuous
version of the multinomial logit choice model. As a result, the probability
distribution of man $m$ choosing a woman with attributes $y$ is given by its
density of probability%
\begin{equation}
\pi _{Y|X}\left( y|x\right) =\frac{\exp \frac{U\left( x,y\right) }{\sigma /2}%
}{\int_{\mathcal{Y}}\exp \left( \frac{U\left( x,y^{\prime }\right) }{\sigma
/2}\right) dy^{\prime }}  \label{condx}
\end{equation}%
which is clearly the extension of the logit formalism to the continuous
choice setting. Similarly, the utility of a woman $w$ with attributes $%
y_{w}=y$ who marries a man with attributes $x$ is 
\begin{equation}
V\left( x_{l}^{w},y\right) +\frac{\sigma }{2}\eta _{l}^{w},  \label{utW}
\end{equation}%
where $V\left( x,y\right) $ is the systematic part of the utility, and $%
\{\left( x_{l}^{w},\eta _{l}^{w}\right) ,l\in \mathbb{N}\}$ are the points
of a Poisson process on $\mathcal{X}\times \mathbb{R}$ of intensity $%
dx\times e^{-\eta }d\eta $, so that the probability distribution of woman $w$
choosing a man with attributes $x$ is given by its density of probability 
\begin{equation}
\pi _{X|Y}\left( x|y\right) =\frac{\exp \frac{V\left( x,y\right) }{\sigma /2}%
}{\int_{\mathcal{X}}\exp \left( \frac{V\left( x^{\prime },y\right) }{\sigma
/2}\right) dx^{\prime }}.  \label{condy}
\end{equation}%
\label{page:CriticismLogit}The continuous logit framework inherits the
structural assumptions of the discrete multinomial logit model. In
particular, the \emph{independence} property, which implies that the
sympathy shock for women whose attributes are in a small set around $y\ $is
perfectly uncorrelated with the sympathy shock for women whose attributes
are in a small set around $y^{\prime }\neq y$. Hence the logit framework
(continuous or discrete) does not allow for a systematic sympathy shock,
i.e. correlated sympathy shocks across observables. In the example where the
attribute of interest is height, it may be desirable to accommodate a random
sympathy shock for height (some men prefer taller women, some prefer shorter
women, regardless of their own observable attributes). We conjecture,
however, that if the amount of variation of the unobserved heterogeneity is
small, the misspecification of the sympathy shocks has only a minor impact
on the market outcome and the identification of the joint utility.

Taking the logarithm of Equations (\ref{condx}) and (\ref{condy})
respectively yields $U-a=(\sigma /2)\log \pi $ and $V-b=(\sigma /2)\log \pi $%
, where%
\begin{equation}
a\left( x\right) =\frac{\sigma }{2}\log \int_{\mathcal{Y}}\frac{e^{\frac{%
U\left( x,y^{\prime }\right) }{\sigma /2}}}{f\left( x\right) }dy^{\prime }%
\text{ and }b\left( y\right) =\frac{\sigma }{2}\log \int_{\mathcal{X}}\frac{%
e^{\frac{V\left( x^{\prime },y\right) }{\sigma /2}}}{g\left( y\right) }%
dx^{\prime },  \label{defaAndb}
\end{equation}%
and since $\Phi =U+V$, one obtains by summation%
\begin{equation}
\log \pi \left( x,y\right) =\frac{\Phi \left( x,y\right) -a\left( x\right)
-b\left( y\right) }{\sigma }.  \label{Schrodinger}
\end{equation}

We formalize this result in Theorem \ref{thm:continuous-CS}, which extends
Galichon and Salani\'{e} (2010) to the continuous case.

\begin{theorem}
\label{thm:continuous-CS}Under the assumptions stated above, the following
holds:

(i) The equilibrium matching $\pi $ maximizes the social gain%
\begin{equation}
\max_{\pi \in \mathcal{M}\left( P,Q\right) }\iint_{\mathcal{X}\times 
\mathcal{Y}}\Phi \left( x,y\right) \pi \left( x,y\right) dxdy-\sigma \iint_{%
\mathcal{X}\times \mathcal{Y}}\log \pi \left( x,y\right) \pi \left(
x,y\right) dxdy.  \label{eq:socialWelfare}
\end{equation}

(ii) In equilibrium, for any $x\in \mathcal{X},y\in \mathcal{Y}$%
\begin{equation}
\pi \left( x,y\right) =\exp \left( \frac{\Phi \left( x,y\right) -a\left(
x\right) -b\left( y\right) }{\sigma }\right)  \label{identEq}
\end{equation}%
where the potentials $a\left( x\right) $ and $b\left( y\right) $ are
determined such that $\pi \in \mathcal{M}\left( P,Q\right) $. They exist and
are uniquely determined up to a constant.

(iii) A man $m$ of attributes $x$ who marries a woman $k^{\ast }$ from his
set of acquaintances obtains utility 
\begin{equation}
U\left( x,y_{k^{\ast }}^{m}\right) +\frac{\sigma }{2}\varepsilon _{k^{\ast
}}^{m}=\max_{k}\left( U\left( x,y_{k}^{m}\right) +\frac{\sigma }{2}%
\varepsilon _{k}^{m}\right)  \label{Utildeeq}
\end{equation}%
where 
\begin{equation}
U\left( x,y\right) =\frac{\Phi \left( x,y\right) +a\left( x\right) -b\left(
y\right) }{2}.  \label{Ueq}
\end{equation}

Similarly, a woman $w$ of attributes $y$ who marries man $l^{\ast }$ from
her set of acquaintances obtains utility%
\begin{equation}
V\left( x_{l^{\ast }}^{w},y\right) +\frac{\sigma }{2}\eta _{l^{\ast
}}^{w}=\max_{l}\left( V\left( x_{l}^{w},y\right) +\frac{\sigma }{2}\eta
_{l}^{w}\right)  \label{Vtildeeq}
\end{equation}%
where%
\begin{equation}
V\left( x,y\right) =\frac{\Phi \left( x,y\right) -a\left( x\right) +b\left(
y\right) }{2}.  \label{Veq}
\end{equation}
\end{theorem}

As in Galichon and Salani\'{e} (2010; 2013), and independently, Decker et
al. (2013), part (i) of this result expresses the fact that the equilibrium
matching reflects a trade-off between sorting on the observed attributes
(which tends to maximize the term $\int \Phi \left( x,y\right) \pi \left(
x,y\right) dxdy$), and sorting on the unobserved attributes (which in turn
tends to maximize the entropic term $\int \log \pi \left( x,y\right) \pi
\left( x,y\right) dxdy$). The second term will therefore pull\ the solution
towards the random matching, where partners are randomly assigned; the
parameter $\sigma $, which captures the intensity of the unobserved
heterogeneity, measures the intensity of this trade-off. The smaller the $%
\sigma $ (i.e. the less unobserved heterogeneity in the model), the closer
the solution will be to the solution without heterogeneity. On the contrary,
the higher the $\sigma $, the larger the probabilistic independence between
the observed attributes of men and women. As an illustration, we consider
the simple toy example below, in which this phenomenon is explicit.

\begin{example}
\label{ex:QuadraticNormal}When $P$ and $Q$ are the standard univariate
Gaussian distribution $\mathcal{N}\left( 0,1\right) $, and $\Phi \left(
x,y\right) =-\frac{1}{2}\left( x-y\right) ^{2}$, the equilibrium matching $%
\pi $ is such that $\pi _{Y|X}\left( y|x\right) =\mathcal{N}\left(
tx,1-t^{2}\right) $, where $t=\sqrt{\frac{\sigma ^{2}}{4}+1}-\frac{\sigma }{2%
}$. Hence, $\sigma =0$ implies $t=1$, in which case $Y=X$ (sorting
predominates and we have positive assortative matching), while $\sigma
\rightarrow \infty $ implies $t\rightarrow 0$, in the limit of which $Y$
becomes independent from $X$ (unobserved heterogeneity predominates, there
is no sorting on observables). Closed-form formulae can also be provided in
the multivariate case when $P$ and $Q$ are Gaussian and $\Phi $ is
quadratic, see Bojilov and Galichon (2013).
\end{example}

Part (ii) of Theorem~\ref{thm:continuous-CS} is an expression of the first
order optimality conditions. The program is an infinite dimensional linear
programming problem where $a\left( x\right) $ and $b\left( y\right) $ are
the Lagrange multipliers corresponding to the constraints $\int \pi \left(
x,y\right) dy=f\left( x\right) $ and $\int \pi \left( x,y\right) dx=g\left(
y\right) $ respectively. Equation (\ref{identEq}), or more precisely its
logarithmic transform Equation (\ref{Schrodinger}), will be the basis of our
estimation strategy. Together with the constraint $\pi \in \mathcal{M}\left(
P,Q\right) $, this equation provides a nonlinear system of equations in $%
a\left( .\right) $ and $b\left( .\right) $. In the applied mathematical
literature it is known as the \emph{Schr\"{o}dinger-Bernstein equation}, or
more commonly as the \emph{Schr\"{o}dinger problem}. Existence and
uniqueness (up to a constant) are well studied under very general conditions
on $P$ and $Q$, see for instance R\"{u}schendorf and Thomsen (1993) and
references therein. \ An efficient algorithm for the numerical determination
of the solution based on a fixed point idea is studied in R\"{u}schendorf
(1995). For completeness it is further explained in Appendix~\ref%
{app:computation}.

Part (iii) of Theorem~\ref{thm:continuous-CS} explains how the joint utility
is shared at equilibrium. Unsurprisingly, just as in Choo and Siow (2006)
and the ensuing literature, individuals do not transfer their sympathy shock
at equilibrium, which is expressed by (\ref{Utildeeq}) and (\ref{Vtildeeq}).
Expressions (\ref{Ueq}) and (\ref{Veq}) provide the formulae for the
systematic part of the utility. As previously noted, $a\left( x\right) $ and 
$b\left( y\right) $ are the Lagrange multipliers of the scarcity constraints
of men's observable attributes $x$ and women's attributes $y$. Hence a
higher $a\left( x\right) $ shall imply a higher relative scarcity for $x$,
and therefore a greater prospect for utility extraction.

\textbf{Identification}. \label{page:ParIdentification}From an
identification perspective, note that equations (\ref{condx}) and (\ref%
{condy}) imply that the observation of $\pi \left( x,y\right) $ leads to
identification of $U\left( x,y\right) $ up to an additive term $c\left(
x\right) $, and similarly, $V\left( x,y\right) $ up to an additive term $%
d\left( y\right) $ by%
\begin{eqnarray*}
U\left( x,y\right) &=&\frac{\sigma }{2}\left( \log \pi _{Y|X}\left(
y|x\right) +c\left( x\right) \right) \text{ and }V\left( x,y\right) =\frac{%
\sigma }{2}\left( \log \pi _{X|Y}\left( x|y\right) +d\left( y\right) \right)
\\
\text{and thus} &&\text{ }\Phi \left( x,y\right) =\frac{\sigma }{2}\left(
\log \pi _{Y|X}\left( y|x\right) +\log \pi _{X|Y}\left( x|y\right) +c\left(
x\right) +d\left( y\right) \right)
\end{eqnarray*}

As a result, $\Phi \left( x,y\right) $ is identified up to a separatively
additive function since we restrict our attention to the matched population%
\footnote{%
Appendix~\ref{app:Singles} explains how these results are extended when
singles are observed.}. Since $\Phi \left( x,y\right) $ yields the same
equilibrium matching as $\Phi \left( x,y\right) +c\left( x\right) +d\left(
y\right) $, the identified quantity is actually the cross-derivative $%
\partial ^{2}\Phi \left( x,y\right) /\partial x\partial y$, while neither $%
\partial \Phi \left( x,y\right) /\partial x$ nor $\partial \Phi \left(
x,y\right) /\partial y$ can be identified, nor can their signs be
identified, either. To illustrate, assume that there is only one
dimension--education. It may be that men and women who are more educated
generate more utility, which we call \textquotedblleft absolute
attractiveness,\textquotedblright\ and which translates into $\partial \Phi
\left( x,y\right) /\partial x\geq 0$ and $\partial \Phi \left( x,y\right)
/\partial y\geq 0$. However, this is not identifiable in our model, because
models where the joint utility is $\Phi \left( x,y\right) $ are
observationally undistinguishable from models where the joint utility is $%
\Phi \left( x,y\right) +c\left( x\right) +d\left( y\right) $, and the terms $%
c$ and $d$ might be strongly negatively correlated with education. Instead,
the present framework allows us to determine whether education is mutually
attractive in the sense that $\partial ^{2}\Phi \left( x,y\right) /\partial
x\partial y\geq 0$, meaning not only that highly educated men and women
attract each other, but also that lower educated men and women attract each
other. Hence, our model allows us to measure the strength of \emph{mutual
attractiveness} (or assortativeness) on various dimensions, but not \emph{%
absolute attractiveness}.

\newpage

\section{Parametric estimation\label{sec:param}}

\subsection{Specification of the matching utility\label{par:specif}}

\label{justifContinuousChooSiow3}In this section, we shall specify a
parametric form for the joint utility function, the estimation of which
shall be discussed in the next section. While Choo and Siow's estimator is
fully nonparametric, the fact that the variables under study are continuous
reinforces the need for a parametric estimator. For the purpose of this
discussion, we shall look back at the illustrative example from the
introduction where only both partners' heights are observed. The Choo and
Siow analysis provides a nonparametric estimator for the joint utility $\Phi
\left( x,y\right) $ of a match between a man of height $x$ and a woman of
height $y$. If heights were to be rounded at the nearest inch and
individuals' heights in inches ranged, say, from 50 to 90, then the
dimension of vector $\Phi \left( x,y\right) $ would be $40\times 40=1600$.
Note that this would worsen significantly if several characteristics were
observed. But even in the single-dimensional case, there would be a serious
missing data problem, since the odds that one would observe data for every
pair of heights are virtually zero. Moreover, even if one were lucky enough
to obtain the full nonparametric estimator of $\Phi \left( x,y\right) $, one
would have to heavily process this information before being able to draw any
stylized conclusion. This simple example highlights the need for a
parametric estimation when considering continuous variables.

Throughout the rest of the paper, we shall assume a quadratic
parametrization of $\Phi $: for $A$ a $d_{x}\times d_{y}$ matrix, we take%
\begin{equation*}
\Phi _{A}\left( x,y\right) =x^{\prime }Ay,
\end{equation*}%
where we call matrix $A$ the \emph{affinity matrix}. One has%
\begin{equation*}
A_{ij}=\frac{\partial ^{2}\Phi \left( x,y\right) }{\partial x_{i}\partial
y_{j}}.
\end{equation*}%
The parameter $A_{ij}$ accounts for the strength of mutual attractiveness
(which can be positive or negative) between dimensions $x_{i}$ and $y_{j}$.
It measures how the (marginal) gain in joint utility from increasing the
man's $i^{th}$ attribute evolves as the woman's $j^{th}$ attribute
increases. It captures the intensity of the complementarity/substitutability
between attribute $x_{i}$ of man $x$ and attribute $y_{j}$ of woman $y$ in
the joint utility.

\label{page:SimplestComplementarity}Two comments about this parametric
choice are noteworthy. First, this parametric choice is arguably the
simplest one which captures nontrivial complementarities between any pair of
attributes. If $A_{ij}>0$, $x_{i}$ and $y_{j}$ are complements, and (all
things else being equal) high $x_{i}$ tend to match with high $y_{j}$. It
reflects positive assortative matching\ across men's $i$-th attribute and
women's $j$-th attribute. For instance, the level of education of one of the
partners may be complementary with the risk aversion of the other partner.
On the contrary, if $A_{ij}<0$, then $x_{i}$ and $y_{j}$ are substitutes,
there is negative assortative matching\ between $x_{i}$ and $y_{j}$. Note
that attributes $x$ and $y$ should not be interpreted as an absolute quality
(where a greater value of $x_{i}$, the $i$-th dimension of $x$, would be
more socially desirable than a smaller value of $x_{i}$). In fact, the model
is observationally undistinguishable from a model where $x$ is changed into $%
-x$ and $y$ is changed into $-y$.

\label{pageCommpletion}Second, this quadratic setting where $\Phi $ is
bilinear in $x$ and $y$ is less restrictive than it seems and can be
extended to the case when{\ the various observed attributes have nonlinear
contributions to the joint utility\footnote{%
We thank a Referee for pointing this out.}. For instance it may be plausible
that extraverted men are indifferent about the education and the height of
their wives, but that if a woman is tall, then men prefer her with a higher
education. Our setting can easily be extended to incorporate such nonlinear
effects. We assume no restrictions on the attributes that enter }$x$ and $y$%
, so that the observables {can be enriched by the addition of nonlinear
functions of them}, i.e. adding $x_{i}^{2}$, $x_{i}^{3}$ etc. and $%
x_{i}x_{j} $ as observable attributes for men and similarly for women. This
will allow $\Phi \left( x,y\right) $ to be any polynomial function of $x$
and $y$. Thus, our setting can easily incorporate any utility function which
is a polynomial expression of the observable attributes.

\subsection{Inference\label{par:inference}}

We turn to the estimation of the affinity matrix $A$. The technique we apply
here was introduced by Galichon and Salani\'{e} (2010); we discuss this
extension to the continuous case. By taking the cross-derivative of
Equation~(\ref{Schrodinger}), one has 
\begin{equation}
\frac{A_{ij}}{\sigma }=\frac{\partial ^{2}\log \pi \left( x,y\right) }{%
\partial x_{i}\partial y_{j}}.  \label{IdentA}
\end{equation}

A seemingly natural procedure would consist in estimating $\pi $
nonparametrically, and obtaining $A$ from the cross derivatives with respect
to $x_{i}$ and $y_{j}$. While feasible, this procedure faces a number of
issues both in theory and in practice. First, it requires a nonparametric
estimation of the second derivatives of the loglikelihood, which is quite
challenging: the \textquotedblleft curse of
dimensionality\textquotedblright\ would fully apply\footnote{%
In our application, both $x$ and $y$ have 10 dimensions, so $\left(
x,y\right) $ is of dimension 20.}\label{page:rkOnCurseOfDim}. Second, since
equation (\ref{IdentA}) is valid at any point $\left( x,y\right) $, this
equation is an over-identifying restriction to the estimation of $A$. The
right hand side of (\ref{IdentA}) depends on $\left( x,y\right) $, while the
left hand side does not. One may certainly take some averaging of the right
hand side of Equation (\ref{IdentA}), but it is not quite obvious how to
weigh each point optimally, and it would only partially offset the problems
stemming from the curse of dimensionality. As a result, this procedure will
be statistically inefficient.

\label{page:description}Instead, we prefer to resort to a moment matching
procedure, which is relatively simple while achieving asymptotic statistical
efficiency as shown in Theorem~\ref{thm:Hessian} below. Let us provide
intuition for this method. Each value of the matrix $A$ yields an
equilibrium matching distribution, which we denote $\pi ^{A}\left(
x,y\right) $. As argued in Appendix~\ref{app:computation}, $\pi ^{A}$ can be
computed efficiently using a fixed point method. Recall that we have assumed
that the distributions of $X$ and $Y$ have zero mean, and introduce the 
\emph{cross-covariance matrix} 
\begin{equation}
\Sigma _{XY}=\left( \mathbb{E}\left[ X_{i}Y_{j}\right] \right) _{ij}=\mathbb{%
E}\left[ XY^{\prime }\right]  \label{crosscov}
\end{equation}%
which is observed in the data. The idea is to look for the value of $A$ such
that for all $i$ and $j$, the covariances predicted by the model match the
covariances observed in the data, that is 
\begin{equation}
\mathbb{E}_{\pi ^{A}}\left[ X_{i}Y_{j}\right] =\mathbb{E}\left[ X_{i}Y_{j}%
\right] .  \label{momentMatching}
\end{equation}

This yields a map $A\rightarrow (\mathbb{E}_{\pi ^{A}}\left[ X_{i}Y_{j}%
\right] )_{ij}$ that is invertible. The inversion of this map (in order to
estimate $A$) can be formulated as a convex optimization problem, thus
making it easy to solve numerically. To see this, we shall recall that the
equilibrium matching $\pi $ maximizes the social gain%
\begin{equation}
\mathcal{W}_{\sigma }\left( A\right) :=\max_{\pi \in \mathcal{M}\left(
P,Q\right) }\mathbb{E}_{\pi }\left[ X^{\prime }AY\right] -\sigma \mathbb{E}%
_{\pi }\left[ \ln \pi \left( X,Y\right) \right] ,  \label{socialGain}
\end{equation}%
and we see that models with parameters $\left( A,\sigma \right) $ and models
with parameters $\left( A/\sigma ,1\right) $ are observationally equivalent,
which translates mathematically into positive homogeneity $\mathcal{W}%
_{\sigma }\left( A\right) =\sigma \mathcal{W}_{1}\left( A/\sigma \right) $.
By the envelope theorem, the predicted covariance between $X_{i}$ and $Y_{j}$
coincides with the partial derivative of $\mathcal{W}_{\sigma }$ with
respect to $A_{ij}$, that is%
\begin{equation}
\mathbb{E}_{\pi ^{A}}\left[ X_{i}Y_{j}\right] =\frac{\partial \mathcal{W}%
_{\sigma }}{\partial A_{ij}}\left( A\right) =\frac{\partial \mathcal{W}_{1}}{%
\partial A_{ij}}\left( A/\sigma \right) ,  \label{envThm}
\end{equation}%
which implies that, upon normalization $\sigma =1$, the map $A\rightarrow (%
\mathbb{E}_{\pi ^{A}}\left[ X_{i}Y_{j}\right] )_{ij}$ is invertible since $%
\mathcal{W}_{1}$ is strictly convex (see Lemma~\ref{lem:strictConv}). This
led Galichon and Salani\'{e} (2013) to conclude, in a setting with discrete
observable attributes, that $B=A/\sigma $ is identified as a solution to the
following convex optimization program%
\begin{equation}
\min_{B\in \mathcal{M}_{d_{x}d_{y}}\left( \mathbb{R}\right) }\left\{ 
\mathcal{W}_{1}\left( B\right) -\sum_{ij}B_{ij}\Sigma _{XY}^{ij}\right\}
\label{MomentMatching}
\end{equation}%
whose first-order conditions are precisely $\partial \mathcal{W}%
_{1}(B)/\partial B_{ij}=\Sigma _{XY}^{ij}$, that is $\mathbb{E}_{\pi ^{B}}%
\left[ X_{i}Y_{j}\right] =\Sigma _{XY}^{ij}$. In the present setting with
continuous observable attributes, things work in an identical manner. Since
the model is scale-invariant, only $A/\sigma $ is identified and we
normalize $A$ so that $\left\Vert A\right\Vert =1$, where $\left\Vert
A\right\Vert =(\sum_{ij}A_{ij}^{2})^{1/2}$. $A$ and $\sigma $ are then
obtained by $A=B/\left\Vert B\right\Vert $ and $\sigma =1/\left\Vert
B\right\Vert $. Let us denote $A^{XY}$ the (unique) solution to this
problem, which will be our estimator of the affinity matrix $A$. Affinity
matrix $A^{XY}$ is \textquotedblleft dual\textquotedblright\ to
cross-covariance matrix $\Sigma _{XY}$ in the sense that there is a
one-to-one correspondence between them by Equation (\ref{envThm}). However,
the former has a structural interpretation: it measures the strength of the
interactions between pairs of attributes.

\label{page:usefulStructural}At this point, it is worth commenting on the
relevance of the structural approach. Indeed, it does not suffice to just
look at the variance-covariance matrix inside matches to infer the sign of
complementarities, as illustrated on the following example. Imagine two
observed characteristics, where the first dimension is education and the
second dimension is risk aversion. Suppose we observe positive correlation
in partners' educations and in partners' risk aversions (i.e., $\Sigma
_{11}>0$ and $\Sigma _{22}>0$). One might naively infer that there is
positive complementarity both in education and in risk aversion (i.e., $%
A_{11}>0$ and $A_{22}>0$). However, this is not necessarily the case; there
could actually be negative complementarity in risk aversion $(A_{22}<0$),
but positive association between individuals' education and risk aversion,
if positive complementarity in education ($A_{11}>0$) dominates the negative
complementarity in risk aversion, thus leading to positive correlation in
risk aversions inside matches. The structural approach allows to avoid this
misinterpretation by allowing to control for the marginal distributions
(e.g. control for the fact that there is positive association between
individuals' education and risk aversion).

Once the affinity matrix $A^{XY}$ has been estimated, two questions arise.
First, what is the rank of $A^{XY}$? This question is of importance since
one would like to know the number of dimensions of $x$ and $y$ on which
sorting occurs. Second, how can we construct \textquotedblleft indices of
mutual attractiveness\textquotedblright\ such that each pair of indices for
men and women explains a mutually exclusive part of the matching utility?
Many studies resort to a technique called \textquotedblleft Canonical
Correlation,\textquotedblright\ which essentially relies on a singular value
decomposition of $\Sigma ^{XY}$. In Dupuy and Galichon (2012), we argue that
this technique is not well suited for studying assortative matching, and
that the resulting procedure is inconsistent. Instead, in Section \ref%
{sec:Saliency}, we propose a method we call \textquotedblleft Saliency
Analysis\textquotedblright\ in order to accurately answer these two
questions. This method is essentially based on the singular value
decomposition of the affinity matrix $A^{XY}$ (instead of $\Sigma ^{XY}$ as
in Canonical Correlation). Testing the rank of the affinity matrix is
equivalent to testing the number of (potentially multiple) singular values
different from $0$. Performing this decomposition allows one to construct
the indices of mutual attractiveness\ that each explain a separate share of
the joint utility.

\newpage

\section{Saliency Analysis\label{sec:Saliency}}

In this section we set out to determine the rank of the affinity matrix $%
A^{XY}$, and the principal dimensions in which it operates. For this, we
introduce and describe a novel technique we call \emph{Saliency Analysis},
which is similar in spirit to Canonical Correlation but does not suffer the
pitfalls of the latter. Instead of performing a singular value decomposition
of the (renormalized) cross-covariance matrix $\Sigma _{XY}$, we shall
perform a singular value decomposition of the affinity matrix $A^{XY}$,
properly renormalized. This idea is similar in spirit to the proposal of
Heckman (2007), who interprets the assignment matrix as a sum of
Cobb-Douglas technologies using a singular value decomposition in order to
refine bounds on wages.

\label{page:DescribeRescale}Recall that we have defined the cross-covariance
matrix $\Sigma _{XY}=\mathbb{E}_{\pi }\left[ XY^{\prime }\right] $, and let
us introduce $S_{X}$ and $S_{Y}$ the diagonal matrices whose diagonal terms
are respectively the variances of the $X_{i}$ and the $Y_{j}$, that is 
\begin{equation*}
S_{X}=diag\left( var(X_{i}),~i=1,...,d_{x}\right) ,~S_{Y}=diag\left(
var(Y_{j}),~j=1,...,d_{y}\right) .
\end{equation*}

We shall work with the rescaled attributes $S_{X}^{-1/2}X$ and $%
S_{Y}^{-1/2}Y $, whose entries each have unit variance. By Lemma~\ref%
{lem:rescale2} in Appendix~\ref{sec:proofThmHessian}, the affinity matrix
between the rescaled attributes $S_{X}^{-1/2}X$ and $S_{Y}^{-1/2}Y$ is%
\begin{equation}
\Theta =S_{X}^{1/2}A^{XY}S_{Y}^{1/2},  \notag
\end{equation}%
for which a singular value decomposition of $\Theta $ yields%
\begin{equation*}
\Theta =U^{\prime }\Lambda V,
\end{equation*}%
where $\Lambda $ is a diagonal matrix with nonincreasing elements $\left(
\lambda _{1},...,\lambda _{d}\right) $, $d=\min \left( d_{x},d_{y}\right) $
and $U$ and $V$ are orthogonal matrices. Define the vectors of \emph{indices
of mutual attractiveness} 
\begin{equation*}
\tilde{X}=US_{X}^{-1/2}X\text{ and }\tilde{Y}=VS_{Y}^{-1/2}Y,
\end{equation*}%
where each index is a weighted sum of the observed attributes. Let $A^{%
\tilde{X}\tilde{Y}}$ be the affinity matrix on the rescaled vectors of
characteristics $\tilde{X}$ and $\tilde{Y}$. From Lemma \ref{lem:rescale2},
it follows that $A^{\tilde{X}\tilde{Y}}=\Lambda $, and as a result 
\begin{equation*}
\Phi _{A}\left( x,y\right)
=\sum_{i=1}^{d_{x}}\sum_{j=1}^{d_{y}}A_{ij}x_{i}y_{j}=\sum_{i=1}^{d}\lambda
_{i}\tilde{x}_{i}\tilde{y}_{i}.
\end{equation*}

\label{page:interpretationXtilde}Hence, the new indices $\tilde{x}$ and $%
\tilde{y}$ are such that $\tilde{x}_{i}$ and $\tilde{y}_{j}$ are complements
for $i=j$, and neither complements, nor substitutes if $i\neq j$. In other
words, there is positive assortative matching between $\tilde{x}_{i}$ and $%
\tilde{y}_{j}$ for $i=j$, and no assortativeness for $i\neq j$. This
justifies the choice of terminology: $\tilde{x}_{i}$ and $\tilde{y}_{i}$ are
\textquotedblleft mutually attractive\textquotedblright\ because they are
complementary with each other and only with each other. All things being
equal, a man with a higher $\tilde{x}_{i}$ tends to match with a woman with
a higher $\tilde{y}_{i}$.

The weights of each index of mutual attractiveness constructed by Saliency
Analysis are given by the associated row of $US_{X}^{-1/2}$ for men and $%
VS_{Y}^{-1/2}$ for women. The value $\lambda _{i}/(\sum_{i}\lambda _{i})$
indicates the share of the observable matching utility of couples explained
by the $i^{th}$ pair of indices. The fact that $U$ and $V$ are orthogonal
implies strong restrictions on how $\tilde{x}$ and $\tilde{y}$ are obtained
from $S_{X}^{-1/2}x$ and $S_{Y}^{-1/2}y$. In particular, this mapping
preserves distances between points; that is, the distance between $\tilde{x}$
and $\tilde{x}^{\prime }$ is equal to the distance between $S_{X}^{-1/2}x$
and $S_{X}^{-1/2}x^{\prime }$.

We observe that in contrast with Canonical Correlation analysis, a
convenient feature of Saliency Analysis is that the results do not change
when the attributes are measured using different measurement units, as
expressed in Lemma~\ref{lem:invariance}. For instance, if the partner's
heights are measured in feet rather than in meters, the outcome of Saliency
Analysis does not change.

For illustrative purposes, we give a stylized example of how Saliency
Analysis operates in a simple two-dimensional situation.

\begin{example}
\label{ex:SVD2D}Assume that there are two dimensions on each side of the
market, and that $S_{X}=S_{Y}=Id$, so that $\Theta =A$. Suppose that%
\begin{equation}
A=%
\begin{pmatrix}
0 & 4 \\ 
-1 & 0%
\end{pmatrix}%
.  \label{exAffinity}
\end{equation}%
Then the singular value decomposition of $A$ is $A=U^{\prime }\Lambda V$,
where%
\begin{equation*}
U=%
\begin{pmatrix}
1 & 0 \\ 
0 & 1%
\end{pmatrix}%
\text{, }\Lambda =%
\begin{pmatrix}
4 & 0 \\ 
0 & 1%
\end{pmatrix}%
\text{ and }V=%
\begin{pmatrix}
0 & 1 \\ 
-1 & 0%
\end{pmatrix}%
\end{equation*}%
The economic interpretation of this simple example is the following: if the
joint utility is given by $\Phi \left( x,y\right) =4x_{1}y_{2}-x_{2}y_{1}$,
then the indices of mutual attractiveness should be given by $\tilde{x}%
_{1}=x_{1}$, $\tilde{x}_{2}=x_{2}$ and $\tilde{y}_{1}=y_{2}$, $\tilde{y}%
_{2}=-y_{1}$. One has $\Phi \left( \tilde{x},\tilde{y}\right) =4\tilde{x}_{1}%
\tilde{y}_{1}+\tilde{x}_{2}\tilde{y}_{2}$. The vectors $\tilde{x}=\left(
x_{1},x_{2}\right) $ and $\tilde{y}=\left( y_{2},-y_{1}\right) $ can be
interpreted as indices of mutual attractiveness, meaning that there is
sorting between $\tilde{x}_{1}=x_{1}$ and $\tilde{y}_{1}=y_{2}$, and between 
$\tilde{x}_{2}=x_{2}$ and $\tilde{y}_{2}=-y_{1}$. If one was willing to
approximate the model by a one-dimensional sorting model, then Saliency
Analysis advocates to keep $\tilde{x}_{1}=x_{1}$ as a proxy for the
attributes of men and $\tilde{y}_{1}=y_{2}$ as a proxy for the attributes of
women. In this case the joint utility is approximated by $\Phi \left( \tilde{%
x},\tilde{y}\right) =4\tilde{x}_{1}\tilde{y}_{1}$.
\end{example}

Example~\ref{ex:SVD2D} is the occasion to compare Singular Value
Decomposition to another matrix decomposition, the Eigenvalue Decomposition,
which economists may be more familiar with. Eigenvalue decomposition
consists in writing, whenever possible, a square matrix $M$ as $M=R\Lambda
R^{-1}$, with $\Lambda $ diagonal and $R$ invertible. In the context of
Saliency Analysis, this decomposition cannot be performed on $A$ as $A$ is
not necessarily a square matrix; further, as soon as $A$ is not symmetric,
this decomposition does not necessarily exist. In particular, when $A$ is
given by (\ref{exAffinity}), it does not exist since $A$ has no real
eigenvalue\footnote{%
However, the singular values of $A$ can be interpreted as eigenvalues of a
larger matrix. Indeed, letting $H$ be the constant Hessian matrix of the map 
$2\Phi $, then $H$ is a symmetric matrix of size $\left( d_{x}+d_{y}\right) $
written blockwise with two zero blocks on its diagonal and $A$ and $%
A^{\prime }$ as off-diagonal blocks, and the eigenvalues of $H$ are plus and
minus the singular values of $A$ (see Horn and Johnson, 1991, p. 135).}.

As Example~\ref{ex:SVD2D} also illustrates, the observation of vector $%
\Lambda $ will allow one to draw conclusions about the multivariate nature
of the sorting, and on the number of dimensions on which the sorting occurs.
In particular, testing for multidimensional sorting versus unidimensional
sorting is equivalent to testing whether at least two singular values $%
\lambda _{i}$ are significantly larger than 0, as we shall elaborate in the
next section.

\newpage

\section{Inferring the number of sorting dimensions\protect\footnote{%
In a discussion with one of the authors, Jim Heckman suggested the intuition
of the approach proposed in this paper to test for multidimensional sorting.}
\label{sec:Asympt}}

Assume that a finite sample of size $n$ is observed. For the sake of
readability, dependence in $n$ of the estimators will be dropped from the
notations. The vector of mutual attraction weights estimated on the sample
is denoted $\hat{\Lambda}$, while the vector of mutual attraction weights in
the population is denoted $\Lambda $. Similarly $\hat{A}$ is the estimator
of $A$ (which was denoted $A^{XY}$ in Section \ref{par:inference}, where the
construction of that estimator is described). Let $\hat{S}_{X}$, $\hat{S}%
_{Y} $ and $\hat{\Sigma}_{XY}$ be the sample estimators of $S_{X}$, $S_{Y}$
and $\Sigma _{XY}$, respectively. For a given quantity $M$, we shall denote 
\begin{equation}
\delta M=\hat{M}-M,  \label{eqDefDiff}
\end{equation}%
the difference between the estimator of $M$ and $M$.

Consider two important matrices associated to the large sample properties of
the model. The Fisher Information matrix is defined by%
\begin{equation}
\mathbb{F}_{kl}^{ij}=\mathbb{E}_{\pi }\left[ \frac{\partial \log \pi \left(
X,Y\right) }{\partial A_{ij}}\frac{\partial \log \pi \left( X,Y\right) }{%
\partial A_{kl}}\right] ,  \label{exprF}
\end{equation}%
where we note that the lines of $\mathbb{F}$ are indexed by pairs of
integers ($ij$), just as the columns of $\mathbb{F}$ are indexed by pairs $%
kl $. (This is due to the fact that the parameter to be estimated, $A_{ij}$,
is not a vector but a matrix). Hence $\mathbb{F}$ is a \textquotedblleft
doubly-indexed matrix,\textquotedblright\ which we shall denote using bold
font. Some basic formalism on doubly-indexed matrices is recalled in
Appendix~\ref{app:Proofs}, Section~\ref{sec:proofThmRankTest}.

Our next result expresses the asymptotic distribution of the estimators of $%
A $, $S_{X}$ and $S_{Y}$. It will be the main building block for testing the
rank of the affinity matrix (and the number of sorting dimensions).

\begin{theorem}
\label{thm:Hessian}The following convergence holds in distribution for $%
n\rightarrow +\infty $:%
\begin{equation*}
n^{1/2}\left( \delta A,\delta S_{X},\delta S_{Y}\right) \Longrightarrow 
\mathcal{N}\left( 0,%
\begin{pmatrix}
\mathbb{F}^{-1} & 0 & 0 \\ 
0 & \mathbb{K}_{XX} & \mathbb{K}_{XY} \\ 
0 & \mathbb{K}_{XY}^{\prime } & \mathbb{K}_{YY}%
\end{pmatrix}%
\right)
\end{equation*}%
where $\mathbb{F}$ has been defined in (\ref{exprF}), $\mathbb{K}_{XY}$ is
defined by%
\begin{equation*}
\left( \mathbb{K}_{XY}\right) _{ij}^{kl}=1_{\left\{ i=j,k=l\right\}
}cov_{\pi }\left( X_{i}X_{j},Y_{k}Y_{l}\right)
\end{equation*}%
and we define similarly $\mathbb{K}_{XX}$ and $\mathbb{K}_{YY}$ by%
\begin{equation*}
\left( \mathbb{K}_{XX}\right) _{ij}^{kl}=1_{\left\{ i=j,k=l\right\}
}cov_{\pi }\left( X_{i}X_{j},X_{k}X_{l}\right) \text{ and }\left( \mathbb{K}%
_{YY}\right) _{ij}^{kl}=1_{\left\{ i=j,k=l\right\} }cov_{\pi }\left(
Y_{i}Y_{j},Y_{k}Y_{l}\right) .
\end{equation*}
\end{theorem}

Note that, as shown in Lemma~\ref{lem:hessianW} in Appendix~\ref%
{sec:proofThmHessian}, $\mathbb{F}$ can be evaluated numerically as the
Hessian matrix of $\mathcal{W}_{1}$. Theorem~\ref{thm:Hessian} implies in
particular that the asymptotic variance-covariance matrix of our estimator
of $A$ is the inverse of the Fisher information matrix. As a result, our
estimator attains asymptotic statistical efficiency.

Now denoting $\hat{\Theta}=\hat{S}_{X}\hat{A}\hat{S}_{Y}$ the estimated
counterpart of $\Theta $ whose singular value decomposition is denoted $\hat{%
\Theta}=\hat{U}^{\prime }\hat{\Lambda}\hat{V}$, we show in Appendix~\ref%
{sec:proofThmRankTest} that $\hat{\Theta}$ is asymptotically normal, and we
give an expression for its asymptotic variance-covariance matrix in Lemma~%
\ref{lem:RescaledAffinity}. We use this asymptotic result to test the rank
of the affinity matrix $\Lambda $. Testing the rank of a matrix is an
important issue with a distinguished tradition in econometrics (see e.g.
Robin and Smith, 2000 and references therein). Here, we use results from
Kleibergen and Paap (2006). One wishes to test the null hypothesis $H_{0}$:
the rank of the affinity matrix is equal to $p=1,2,...,d-1$. Following
Kleibergen and Paap, the singular value decomposition $\hat{\Theta}=\hat{U}%
^{\prime }\hat{\Lambda}\hat{V}$ is written blockwise%
\begin{equation*}
\hat{\Theta}=%
\begin{pmatrix}
\hat{U}_{11}^{\prime } & \hat{U}_{21}^{\prime } \\ 
\hat{U}_{12}^{\prime } & \hat{U}_{22}^{\prime }%
\end{pmatrix}%
\begin{pmatrix}
\hat{\Lambda}_{1} & 0 \\ 
0 & \hat{\Lambda}_{2}%
\end{pmatrix}%
\begin{pmatrix}
\hat{V}_{11} & \hat{V}_{12} \\ 
\hat{V}_{21} & \hat{V}_{22}%
\end{pmatrix}%
\end{equation*}%
where the blocks are dimensioned so that $\hat{\Lambda}_{1}$, $\hat{U}%
_{11}^{\prime }$ and $\hat{V}_{11}$ are $p\times p$ square matrices. Define%
\begin{eqnarray*}
\hat{T}_{p} &=&\left( \hat{U}_{22}^{\prime }\hat{U}_{22}\right) ^{-1/2}\hat{U%
}_{22}^{\prime }\hat{\Lambda}_{2}\hat{V}_{22}\left( \hat{V}_{22}^{\prime }%
\hat{V}_{22}\right) ^{-1/2} \\
\hat{A}_{p\perp } &=&%
\begin{pmatrix}
\hat{U}_{21}^{\prime } \\ 
\hat{U}_{22}^{\prime }%
\end{pmatrix}%
\left( \hat{U}_{22}^{\prime }\right) ^{-1}\left( \hat{U}_{22}^{\prime }\hat{U%
}_{22}\right) ^{1/2} \\
\hat{B}_{p\perp } &=&\left( \hat{V}_{22}^{\prime }\hat{V}_{22}\right) ^{1/2}%
\hat{V}_{22}^{-1}%
\begin{pmatrix}
\hat{V}_{21} & \hat{V}_{22}%
\end{pmatrix}%
\end{eqnarray*}%
so that our next result provides a test for the number of sorting dimensions.

\begin{theorem}
\label{thm:rankTest}Under the null hypothesis that the rank of the affinity
matrix is $p$, the quantity\ $n^{1/2}\hat{T}_{p}$ is asymptotically normally
distributed, and the expression of its variance-covariance matrix $\Omega
_{p}$ is given in Appendix~\ref{sec:proofThmRankTest}, formula (\ref%
{exprOmegap}). As a result, the test-statistic%
\begin{equation*}
n\hat{T}_{p}^{\prime }\hat{\Omega}_{p}^{-1}\hat{T}_{p}
\end{equation*}%
converges under the null hypothesis to a $\chi ^{2}\left( \left(
d_{x}-p\right) \left( d_{y}-p\right) \right) $ random variable.
\end{theorem}

\newpage

\section{The data\label{sec:Data}}

\subsection{The dataset}

In this paper, we use the waves 1993-2002 of the DNB Household Survey (DHS)
to estimate preferences in the marriage market. For a thorough description
of the setup and the quality of this data we refer the reader to Nyhus
(1996). This data is a representative panel of the Dutch population with
respect to region, political preference, housing, income, degree of
urbanization, and age of the head of the household among others. The DHS
data was collected via on-line terminal sessions and each participating
family was provided with a PC and a modem if necessary. The panel contains
on average about 2,200 households in each wave.

This data includes three main features that are particularly attractive for
our purposes. First, within each household, all persons aged 16 or over were
interviewed. This implies that the data contains detailed information not
only about the head of the household but about all individuals in the
household. In particular, the data identifies \textquotedblleft
spouses\textquotedblright\ and \textquotedblleft permanent
partners\textquotedblright\ of the head of each household. This information
reveals the nature of the relationship between the various individuals of
each household and allows us to reconstruct \textquotedblleft
couples\textquotedblright .

Second, this data contains very detailed information about individuals. This
rich set of information includes socio-demographic variables such as birth
year and education, as well as variables about the anthropometry of
respondents (height and weight), a self-assessed measure of health, and,
above all, information about personality traits and risk attitude, which are
included in the waves 1993-2002.

Finally, as for most panel data, the DHS data suffers from attrition
problems. The attrition of households is on average 25\% each year, cf. Das
and van Soest (1999) among others. To remedy attrition, refreshment samples
were drawn each year, such that, over the period 1993-2002, about 7,700
distinct households were interviewed at least once. Since the methodology
implemented in this paper relies essentially on the availability of a
cross-section of households, attrition and its remedy is in fact an asset of
this data as it allows us to have access to a rather large pool of potential
couples.

Note that our methodology could be applied on other panel datasets (such as
GSOEP, for instance) that also include supplementary questionnaires enabling
one to construct measures of personality traits and risks aversion together
with socio-demographic and morphological variables. However, the main asset
of the DNB dataset is that it allows us to measure all relevant variables in
a single wave whereas in GSOEP, one would have to use the panel structure to
match measures of BMI (from wave 2008 and 2009) and measures of personality
traits (from waves 2006 and 2007).

\subsection{Variables}

Educational attainment is measured from the respondent's reported highest
level of education achieved. The respondents could choose among 13
categories (7 in the later waves), ranging from primary to university
education. The reduction to 7 categories in the later waves implies that
only three broad educational categories can be consistently constructed. We
coded responses as follows:

\begin{enumerate}
\item Lower [kindergarten, primary, elementary secondary] education,

\item Intermediate [secondary, pre-university, vocational] education,

\item Higher [university] education.
\end{enumerate}

The respondents were also asked about their height and weight. The answers
of these questions allowed us to calculate the Body Mass Index of each
respondent as the weight in kilograms divided by the square of the height
measured in meters. The respondents were also asked to report their general
health. The phrasing of the question was: \textquotedblleft How do you rate
your general health condition on a scale from 1, excellent, to 5,
poor?\textquotedblright . We make use of the panel structure to deal partly
with nonresponses on socio-economic and health variables. When missing
values for height, weight, education, year of birth etc. were encountered,
values reported in adjacent years were imputed. We defined our measure of
health by subtracting the answer to this question to $6$.

In Appendix~\ref{app:Variables}, we recall the methodology of Nyhus and
Webley (2001) which we followed in order to construct five factors of
personality traits. These factors were labeled as:

\begin{itemize}
\item Emotional stability: a high score indicates that the person is less
likely to interpret ordinary situations as threatening, and minor
frustrations as hopelessly difficult,

\item Extraversion (outgoing): a high score indicates that the person is
more likely to need attention and social interaction,

\item Conscientiousness (meticulous): a high score indicates that the person
is more likely to be self-disciplined and plan his/her actions,

\item Agreeableness (flexibility): a high score indicates that the person is
more likely to be pleasant with others and go out of their way to help
others,

\item Autonomy (tough-mindedness): a high score indicates that the person is
more likely to be direct, rough and dominant.
\end{itemize}

\subsection{Couples}

Our definition of a couple is a man and a woman living in the same household
and reporting being either head of the household, spouse of the head or a
permanent partner (not married) of the latter\footnote{%
Note that using the subsample of legally married couples does not affect the
three main results of our analysis mentioned in the abstract. These results
are available from the authors upon request.}. To construct our dataset of
couples, we first pool all the selected waves (1993-2002). We then keep only
those respondents that report being head of the household, spouse of the
head or permanent (not married) partner of the head. This sample contains
roughly 13,000 men and women and identifies about 7,700 unique households.
We then split this sample and create two datasets, one containing women and
one containing men. Each dataset identifies about 6,500 different men and
women. We then merge the men dataset to the women dataset using the
household identifier. We identify 5,500 unique couples while roughly 1,250
men and 1,250 women remain unmatched.

Given the aim of our main analysis, we further restrict our sample to
relatively newly formed couples. In the absence of information about when
couples actually formed, following the literature (see Chiappori et al.
2012, for instance), we select only couples whose wives are younger than 40
years old.

Table \ref{tab0} reports the number of identified young couples and the
number of young couples for which we have complete information on the
various dimensions. For nearly all couples we have information on both
spouses' educational attainment. However, out of 2,897 couples only 1,595
provide complete information on education, height, health and BMI. We lose
another 337 couples for which personality traits are not fully observed.
Another 100 couples are lost when attitude towards risks is additionnally
taken into consideration. Our working dataset therefore contains 1158 young
couples.

Table \ref{tab1} presents summary statistics for men and women. On average,
in our sample, men are 3 years older than women\footnote{%
Note that the mean age at first marriage is relatively high in the
Netherlands (30.1 and 32.8 for women and men respectively, source: United
Nations Economic Commission for Europe, 2010 Statistical Database) compared
to the USA (26.1 and 28.2) which is reflected in the relative high mean age
of men and women in our sample of couples.},\label{page:AgeOfCouples}
slightly more educated, taller by 13 centimeters, have a BMI of 1Kg/m$^{2}$
higher, are less conscientious (meticulous), less extraverted but more
emotionally stable and more risk averse. On average, in our sample, men and
women have similar (good) health and a comparable degree of agreeableness
and autonomy.

Oreffice and Quintana-Domeque (2010) estimate features of the (observed)
matching function between men and women in the marriage market using the
PSID data for the US. Their strategy consists in regressing each attribute
of men on all attributes of women and vice versa. This procedure can easily
be replicated with our data in an attempt to compare features of the
matching function in both datasets (US versus Dutch marriage market).
Interestingly enough, we find very similar results as those obtained by
Oreffice and Quintana-Domeque (2010). For instance, these authors find that
an additional unit in the husband's BMI is associated with a 0.4 additional
unit in the wife's BMI. Using our sample, our estimate is also significant
and of similar magnitude even after controlling for personality traits, i.e.
0.25. Furthermore, they find that an additional inch in the husband's height
is associated with an additional 0.12 inch in the wife's height. Here too,
our estimate is significant and of similar magnitude, i.e. 0.15. Yet,
Oreffice and Quintana-Domeque (2010) find that richer men (higher educated
men in our case) tend to be married with wives of lower BMI (an increase of
10\% in the husband's earnings is associated with a decrease of 0.21 points
in his wife's BMI). In our sample, we find that higher educated men
(interpreting education as permanent income) are matched with women of lower
BMI, i.e. a man with one additional level of education is matched with a
woman whose BMI is 0.56 units lower.

\newpage

\section{Empirical results\label{sec:EmpiricalResults}}

We apply the Saliency Analysis, outlined in the previous section, on our
sample of couples. The procedure requires first to estimate the affinity
matrix $A$. This is done by applying the technique presented in section 3.
The estimation results are reported in Table \ref{tab3}. It is important to
note that the estimates reported in the table are obtained using
standardized attributes rather than the original ones. The main advantage of
using standardized attributes is that the magnitude of the coefficients is
directly comparable across attributes, allowing a direct interpretation in
terms of comparative statics.

The estimates of the affinity matrix reveal four important and remarkable
features:

\begin{enumerate}
\item \textit{On-diagonal}: education is the single most important attribute
in the marriage market. The largest coefficient of the affinity matrix is
indeed observed on the diagonal for education. This coefficient is more than
twice as large as the second largest coefficient obtained on the diagonal
for the variable BMI. Loosely speaking, this means that increasing the
education of both spouses by 1 standard deviation increases the couple's
joint utility by 0.56 units. To achieve a similar increase in utility, the
BMI of both spouses should be increased by 1.63 standard deviations each.

\item \textit{Off-diagonal}: the table clearly indicates the importance of
cross-gender interactions between the various attributes as many
off-diagonal coefficients of the affinity matrix are significantly different
from 0. This implies that important trade-offs take place between the
various attributes. For instance, men's emotional stability interacts
positively with women's conscientiousness, i.e. 0.21. Stated otherwise, this
means that increasing the husband's emotional stability increases the joint
utility of couples whose wives are relatively conscientious. Other examples
are noticeable: men's autonomy interacts negatively with women's
conscientiousness, i.e. $-0.11$ but positively with women's extraversion,
i.e. $0.11$. Conversely, men's agreeableness interacts positively with
women's conscientiousness, i.e. 0.13, but negatively with women's
extraversion, i.e. -0.14.

\item \textit{Asymmetry:} the affinity matrix is not symmetric indicating
that preferences for attributes are not similar for men and women. For
instance, increasing a wife's conscientiousness by 1 standard deviation
increases the joint utility of couples with more agreeable men relatively
more (significant coefficient of magnitude $0.13$) while increasing the
husband's conscientiousness by 1 standard deviation has the same impact on a
couple's joint utility, indifferently of how agreeable his wife is.

\item \textit{Personality traits:} personality traits matter for
preferences, not only directly (terms on the diagonal are significant for
conscientiousness, and risk aversion and of respective magnitude, 0.14 and
0.11) but mainly indirectly through their interactions with other
attributes. For instance, the single most important interaction between
observable attributes of men and women is found between the emotional
stability of husbands and the conscientiousness of women, i.e. 0.21, a
magnitude that matches with the direct effect of BMI. Also, personality
traits interact not only with other personality traits but also with
anthropometry. The emotional stability of men interacts positively with
women's BMI, i.e. 0.12.
\end{enumerate}

Using the estimated affinity matrix, we then proceed to the Saliency
Analysis as introduced in the previous section. This enables us i) to test
whether sorting is unidimensional, i.e. occurs on a single-index and ii) to
construct pairs of indices of mutual attractiveness for men and women.

We first test the dimensionality of the sorting in the marriage market. For $%
p=1$, that is testing against the null hypothesis that sorting occurs on a
single index, we find that $n\hat{T}_{1}^{\prime }\hat{\Omega}_{1}^{-1}\hat{T%
}_{1}=273.45$ which is significant at the 1\% level. This implies that
sorting in the marriage market does not occur on a single index as has been
assumed in most of previous literature. In fact, our test-statistic never
becomes insignificant. Even for $p=9$ we have $n\hat{T}_{9}^{\prime }\hat{%
\Omega}_{9}^{-1}\hat{T}_{9}=13.62$ which is still significant at the 1\%
level. This suggests that the affinity matrix has full rank and that sorting
occurs on at least 10 observed indices. Our results therefore clearly
highlight that sorting in the marriage market is multidimensional and
individuals face important trade-offs between the attributes of their
spouses.

Each pair of indices derived from Saliency Analysis explains a mutually
exclusive part of the total observable matching utility of couples. The
share explained by each of our 10 indices is reported in Table \ref{tab5a}.
The table shows that the share of the first 8 pairs of indices is
significantly different from 0 at the 1\% level.

As for the Principal Component Analysis, the labeling of each dimension is
subjective and becomes increasingly difficult to interpret as one considers
more dimensions. Table \ref{tab5} therefore only contains the 3 pairs of
indices explaining most of the joint utility. Together these 3 pairs of
indices explain about 60\% of the total matching utility. The first pair,
indexed I1, explains about 28\% of the joint utility. These indices load
heavily (in bold weight $\geq$ 0.5) on education and the weights on
education are of similar magnitude for men and women. This confirms that
education plays the most important role in sorting in the marriage market.
However, the second pair of indices, which explains another 17\% of the
joint utility, loads heavily on personality traits (i.e. emotional stability
for men and conscientiousness for women). Personality traits play a strong
role in the sorting in the marriage market. Interestingly enough, while
conscientiousness only matters for the attractiveness of women, emotional
stability only matters for the attractiveness of men. The third pair
explains another 14\% of the joint utility and loads on BMI and extraversion
for women and agreableness for men. This result corroborates Chiappori,
Oreffice and Quintana-Domeque's (2012) finding that BMI is important for the
sorting in the marriage market.

\newpage

\section{Summary and Discussion\label{sec:Conc}}

This paper has introduced a novel technique to test for the dimensionality
of the sorting in the marriage market, and derive indices of mutual
attractiveness, namely Saliency Analysis. This technique is grounded in the
structural equilibrium model of Choo and Siow (2006) which we have extended
to the continuous case in this paper. Indices of mutual attractiveness
derived in Saliency Analysis, in contrast to Canonical Correlation for
instance, have a structural interpretation and are therefore informative
about agents' preferences.

Saliency Analysis has been performed on a dataset of Dutch households
containing information about education, height, BMI, health, attitude
towards risk and five personality traits of both spouses. The empirical
results of this paper reveal two important features of the marriage market.
First, our results clearly show that sorting occurs on multiple indices
rather than just on a single one, as assumed in most of current literature.
This implies that individuals face important trade-offs between the
attributes of their potential spouse. For instance, in the dataset we
studied, more conscientious men prefer more conscientious women (0.14), but
more autonomous men prefer less conscientious women (-0.11). Hence, women
face a trade off between being attractive for more conscientious men and
being attractive for more autonomous men. Similarly, more conscientious
women prefer more agreeable men (0.13) but more extraverted women prefer
less agreeable men (-0.14). Men therefore face a trade-off between being
attractive for more conscientious women and being attractive for more
extraverted women.

Second, personality traits and attitude towards risk matter for the sorting
of spouses in the marriage market. In fact, although education explains the
largest share (28\%) of the observable joint utility of spouses, personality
traits explain a rather large share too (17\%). Interestingly enough,
different traits matter differently for men and women. For instance, women
find emotionally stable men more attractive. Yet, men prefer conscientious
women but are indifferent about the emotional stability of women.

The analysis presented in this paper opens up interesting possibilities for
further research. In particular, our analysis could be applied on other
markets besides the marriage one, such as the market for CEOs. A recent
literature led by Bertrand and Schoar (2003), Falato, Li and Milbourn (2012)
and Custodio, Ferreira and Matos (2013), acknowledges the
multidimensionality of CEO's talent, but assumes that sorting occurs on a
single index. Our setting can then be used to extend the seminal
contributions of Tervi\"{o} (2008) and Gabaix and Landier (2008), who
calibrate a single-dimensional multiplicative sorting model in order to
explain CEO compensation. An important difference in the CEO compensation
literature is that transfers (i.e. salaries) are typically observed, unlike
in the case of the marriage market considered in the present paper. The
observation of the transfers has interesting consequences for
identification. Assume that $x$ is a CEO's vector of characteristics (say,
track record, education, political inclinations, cultural affinities)\ and $%
y $ is a vector of firm's characteristics. Let $\alpha \left( x,y\right) $
be the nonpecuniary utility of CEO $x$ working with firm $y$, and let $%
\gamma \left( x,y\right) $ be the productivity (in monetary units) of CEO $x$
if hired by firm $y$. In the case where transfers are unobserved, only the
joint utility $\Phi =\alpha +\gamma $ is identified. However, in the case
where transfers are observed, it is possible to identify separately $\alpha $
and $\gamma $. Hence when CEO compensation data is available, the results of
the present paper can be easily extended to identify \emph{simultaneously}
the CEO's productivity and his/her nonpecuniary utility for working with a
given firm.

Lastly, we observe that the Poisson process approach which appears in the
framework of this paper may provide the \textquotedblleft missing
link\textquotedblright\ between search models and matching with unobserved
heterogeneity. Indeed, Poisson processes are central to search models, and
the fact that they also play an important role in our model suggests that
they may provide an interesting connection. The key difference, of course,
comes from the fact that in search models, agents are faced with an optimal
stopping problem: agents cannot know what their opportunities will be in
advance, and they cannot retain offers, while in our framework they are
fully aware of all their opportunities from the start. While we briefly
elaborate on the formal connection in Appendix~\ref{app:ContinuousLogit}, we
leave a full exploration of the matter for future work.

\newpage

\appendix

\section{Continuous logit formalism\label{app:ContinuousLogit}}

In this paragraph, we expound the main ideas of Cosslett (1988) and Dagsvik
(1994) who show how to obtain a continuous version of the multinomial logit
model. Assume that $\{\left( y_{k}^{m},\varepsilon _{k}^{m}\right) ,k\in 
\mathbb{N}\}$ are the points of a Poisson point process on $\mathcal{Y}%
\times \mathbb{R}$ of intensity $dy\times e^{-\varepsilon }d\varepsilon $.
We recall that this implies that for $S$ a subset of $\mathcal{Y}\times 
\mathbb{R}$, the probability that man $m$ has no acquaintance in set $S$ is $%
\exp \left( -\int_{S}e^{-\varepsilon }dyd\varepsilon \right) $. From (\ref%
{utMM}), man $m$ chooses woman $k$ among his acquaintances such that his
utility is maximized, that is, man $m$ solves%
\begin{equation*}
\max_{k}\left\{ U\left( x,y_{k}^{m}\right) +\varepsilon _{k}^{m}\right\} .
\end{equation*}%
Letting $Z$ be the value of this maximum, one has for any $c\in \mathbb{R}$ 
\begin{equation*}
\Pr \left( Z\leq c\right) =\Pr \left( U\left( x,y_{k}^{m}\right)
+\varepsilon _{k}^{m}\leq c\text{ }\forall k\right)
\end{equation*}%
which is exactly the probability that the Poisson point process $\left(
y_{k},\varepsilon _{k}^{m}\right) $ has no point in $\left\{ \left(
y,\varepsilon \right) :U\left( x,y\right) +\varepsilon >c\right\} $, thus 
\begin{eqnarray*}
\log \Pr \left( Z\leq c\right) &=&-\iint_{\mathcal{Y}\times \mathbb{R}%
}1\left( U\left( x,y\right) +\varepsilon >c\right) dye^{-\varepsilon
}d\varepsilon =-\int_{\mathcal{Y}}\int_{c-U\left( x,y\right)
}e^{-\varepsilon }d\varepsilon dy \\
&=&-\int_{\mathcal{Y}}e^{-c+U\left( x,y\right) }dy=-\exp \left( -c+\log
\int_{\mathcal{Y}}\exp U\left( x,y\right) dy\right) ,
\end{eqnarray*}%
hence $Z$ is a $\left( \log \int_{\mathcal{Y}}\exp U\left( x,y\right)
dy,1\right) $-Gumbel. In particular, 
\begin{equation*}
\mathbb{E}\left[ \max_{k}\left\{ U\left( x,y_{k}^{m}\right) +\varepsilon
_{k}^{m}\right\} \right] =\log \int_{\mathcal{Y}}\exp U\left( x,y\right) dy
\end{equation*}%
and the choice probabilities are given by their density with respect to the
Lebesgue measure%
\begin{equation*}
\pi \left( y|x\right) =\exp (U\left( x,y\right) )/(\int_{\mathcal{Y}}\exp
U\left( x,y^{\prime }\right) dy^{\prime }).
\end{equation*}

The same logic also implies that $\{\varepsilon _{k}:k\in \mathbb{N\}}$ has
a Gumbel distribution. Indeed, the probability that this Poisson point
process has no element in the set $\left\{ \varepsilon :\varepsilon
>c\right\} $ is equal to%
\begin{equation*}
\exp \left( -\int_{c}^{+\infty }e^{-\varepsilon }d\varepsilon \right) =\exp
\left( -\exp \left( -c\right) \right)
\end{equation*}%
which is equivalent to say that $\Pr \left( \max_{k\in \mathbb{N}%
}\varepsilon _{k}\leq c\right) =\exp \left( -\exp \left( -c\right) \right) $%
. Finally, note that a similar argument would show that $m$ has almost
surely an infinite, though countable, number of acquaintances, as announced.

Note that an interesting connection remains to be explored with the search
literature (Shimer and Smith, 2000, Atakan, 2006). Assume that each man $m$
draws a Poisson sample of acquaintances $\left( y_{k}^{m},\varepsilon
_{k}^{m}\right) $, where $y_{k}^{m}$ is the type of partner of index $k$,
and $\varepsilon _{k}^{m}$ is now the time at which this acquaintance is
met. Assume that it has been agreed that if $x$ matches with $y$, $x$ will
receive utility $U\left( x,y\right) $ out of the joint utility $\Phi \left(
x,y\right) $. In the spirit of Atakan (2006), assume unmatched agents pay a
utility cost equal to $\sigma $ per unit of time while unmatched, such that
if $x$ matches with $y$ at time $\varepsilon $, his utility is $U\left(
x,y\right) -\sigma \varepsilon $. If agents could perfectly foresee their
opportunities (i.e., know the full sample $\left( y_{k}^{m},\varepsilon
_{k}^{m}\right) $ in advance), they would choose opportunity $k$ so as to
maximize the quantity $U\left( x^{m},y_{k}^{m}\right) -\sigma \varepsilon
_{k}^{m}$ exactly as in the present paper. The difference, of course, comes
from the fact that in search models, agents are faced with an optimal
stopping problem: agents cannot know what their opportunities will be in
advance, and they cannot retain offers. At each time $t$ they know only what
opportunities have already been received up to time $t$, and they do not
know about the set of $k$'s such that $\varepsilon _{k}^{m}>t$. This is an
optimal stopping problem with a Poisson process, well-studied in Probability
Theory and Operations Research following seminal work by Elfving (1967). The
basic idea is as follows: there exists a function $\psi :\mathbb{R}%
\rightarrow \mathbb{R}$ such that the partner chosen by $m$ is the first
partner (in terms of meeting time) such that $U\left( x^{m},y_{k}^{m}\right) 
$ exceeds $\psi \left( \varepsilon _{k}^{m}\right) $. $\psi $ can be
characterized as a solution to an Ordinary Differential Equation, and in
some cases, can be expressed analytically.

\newpage

\section{Proofs\label{app:Proofs}}

\subsection{Proof of Theorem \protect\ref{thm:continuous-CS}}

\begin{proof}
(i) The first part of the argument extends Galichon and Salani\'{e} (2010)
to the continuous case; the argument is decomposed in four steps which are
now briefly commented. In Step 1, we shall show that the expression of the
social welfare is given by%
\begin{eqnarray}
&&\min_{U,V}\int_{\mathcal{X}}G_{x}\left( U\left( x,.\right) \right) f\left(
x\right) dx+\int_{\mathcal{Y}}H_{y}\left( V\left( .,y\right) \right) g\left(
y\right) dy  \label{dualGeneral} \\
&&s.t.~U\left( x,y\right) +V\left( x,y\right) \geq \Phi \left( x,y\right) 
\notag
\end{eqnarray}%
where $U\left( x,y\right) $ (resp. $V\left( x,y\right) $) is the share of
the systematic joint utility going to man $x$ (resp. woman $y$), and $%
G_{x}\left( U\right) $ (resp. $H_{y}\left( V\right) $) is the ex-ante
indirect utility of a man of type $x$ (resp. a woman of type $y$), namely%
\begin{equation}
G_{x}\left( U\left( x,.\right) \right) =\mathbb{E}\left[ \max_{k}\left\{
U\left( x,y_{k}^{m}\right) +\frac{\sigma }{2}\varepsilon _{k}^{m}\right\} %
\right] \text{ and }H_{y}\left( V\left( .,y\right) \right) =\mathbb{E}\left[
\max_{l}\left\{ U\left( x_{l}^{w},y\right) +\frac{\sigma }{2}\varepsilon
_{l}^{w}\right\} \right] .  \label{exprGandH}
\end{equation}

Welfare expression (\ref{dualGeneral}) has a straightforward interpretation
in terms of equilibrium. The constraint $U+V\geq \Phi $ is a stability
condition, and the minimization of the sum of the individual ex-ante
indirect utility function expresses the absence of rents.

In step 2, we shall express the dual of variational problem (\ref%
{dualGeneral}) as 
\begin{equation*}
\mathcal{W}=\sup_{\pi \in \mathcal{M}\left( P,Q\right) }\int \Phi d\pi -%
\mathcal{I}\left( \pi \right)
\end{equation*}%
where%
\begin{eqnarray*}
\mathcal{I}\left( \pi \right) &=&\sup_{U}\left( \int_{\mathcal{X}\times 
\mathcal{Y}}U\left( x,y\right) d\pi \left( x,y\right) -\int_{\mathcal{X}%
}G_{x}(U\left( x,.\right) )dP\left( x\right) \right) \\
+ &&\sup_{V}\left( \int_{\mathcal{X}\times \mathcal{Y}}V\left( x,y\right)
d\pi \left( x,y\right) -\int_{\mathcal{Y}}H_{y}\left( V\left( .,y\right)
\right) dQ\left( y\right) \right) .
\end{eqnarray*}

In step 3, we shall show that under the distributional assumptions made on
the heterogeneities, the expression of $\mathcal{I}$ is given by%
\begin{equation*}
\mathcal{I}\left( \pi \right) =\sigma \iint_{\mathcal{X}\times \mathcal{Y}%
}\log \frac{\pi \left( x,y\right) }{\sqrt{f\left( x\right) g\left( y\right) }%
}\pi \left( x,y\right) dxdy
\end{equation*}

In step 4, we shall show that as a result, the social welfare, which is the
value of variational problem (\ref{dualGeneral}), can be expressed up to
irrelevant constants as%
\begin{equation}
\max_{\pi \in \mathcal{M}\left( P,Q\right) }\iint_{\mathcal{X}\times 
\mathcal{Y}}\Phi \left( x,y\right) \pi \left( x,y\right) dxdy-\sigma \iint_{%
\mathcal{X}\times \mathcal{Y}}\log \pi \left( x,y\right) \pi \left(
x,y\right) dxdy  \label{SWF}
\end{equation}%
which will establish (i).

\textbf{Step 1}. Introduce $\varepsilon _{m}\left( .\right) $ a stochastic
process on $\mathcal{Y}$ defined by%
\begin{equation*}
\varepsilon _{m}\left( y\right) =\frac{\sigma }{2}\max_{k}\left\{
\varepsilon _{k}^{m}:y_{k}=y\right\}
\end{equation*}%
if the set $\left\{ k:y_{k}=y\right\} $ is nonempty, $\varepsilon _{m}\left(
y\right) =-\infty $ otherwise. Similarly, introduce $\eta _{w}\left(
x\right) $ a stochastic process on $\mathcal{X}$ defined by%
\begin{equation*}
\eta _{w}\left( x\right) =\frac{\sigma }{2}\max_{l}\left\{ \eta
_{l}^{w}:x_{l}=x\right\}
\end{equation*}%
if the set $\left\{ l:x_{l}=x\right\} $ is nonempty, $\eta _{w}\left(
x\right) =-\infty $ otherwise. By the results of Shapley and Shubik (1972),
extended to the continuous case by Gretsky, Ostroy and Zame (1992), the
equilibrium matching solves the dual transportation problem which expresses
the social welfare 
\begin{equation}
\mathcal{W}=\inf_{u_{m}+v_{w}\geq \Phi \left( x_{m},y_{m}\right)
+\varepsilon _{m}\left( y\right) +\eta _{w}\left( x\right) }\int
u_{m}dm+\int v_{w}dw  \label{Dual}
\end{equation}%
now, the constraint can be rewritten as%
\begin{equation*}
U\left( x,y\right) +V\left( x,y\right) \geq \Phi \left( x,y\right)
\end{equation*}%
where $U$ and $V$ have been defined as%
\begin{equation*}
U\left( x,y\right) =\inf_{m}\left( u_{m}-\varepsilon _{m}\left( y\right)
\right) \text{ and }V\left( x,y\right) =\inf_{w}\left( v_{w}-\eta _{w}\left(
x\right) \right)
\end{equation*}%
which implies that $u_{m}$ and $v_{w}$ can be expressed in $U\left(
x,y\right) $ and $V\left( x,y\right) $ by 
\begin{equation}
u_{m}=\sup_{y\in \mathcal{Y}}\left( U\left( x,y\right) +\varepsilon
_{m}\left( y\right) \right) \text{ and }v_{w}=\sup_{x\in \mathcal{X}}\left(
V\left( x,y\right) +\eta _{w}\left( x\right) \right) .
\label{SurplusTransform}
\end{equation}%
Therefore, replacing $u_{m}$ and $v_{w}$ by their expression in $U$ and $V$,
(\ref{Dual}) rewrites as (\ref{dualGeneral}), with $G_{x}$ and $H_{y}$ given
by (\ref{exprGandH}).

\textbf{Step 2}. Rewrite (\ref{dualGeneral}) as a saddlepoint problem%
\begin{equation*}
\mathcal{W}=\inf_{U,V}\sup_{\pi }\left( 
\begin{array}{c}
\iint_{\mathcal{X}\times \mathcal{Y}}\Phi d\pi \\ 
+\int_{\mathcal{X}}G(U\left( x,.\right) )dP\left( x\right) -\iint_{\mathcal{X%
}\times \mathcal{Y}}Ud\pi \\ 
+\int_{\mathcal{Y}}H\left( V\left( .,y\right) \right) dQ\left( y\right)
-\iint_{\mathcal{X}\times \mathcal{Y}}Vd\pi%
\end{array}%
\right)
\end{equation*}%
or in other words%
\begin{equation*}
\mathcal{W}=\sup_{\pi }\int \Phi d\pi -\mathcal{I}\left( \pi \right)
\end{equation*}%
where%
\begin{eqnarray*}
\mathcal{I}\left( \pi \right) &=&\sup_{U}\left( \iint_{\mathcal{X}\times 
\mathcal{Y}}Ud\pi -\int_{\mathcal{X}}G_{x}(U\left( x,.\right) )dP\left(
x\right) \right) \\
+ &&\sup_{V}\left( \iint_{\mathcal{X}\times \mathcal{Y}}Vd\pi -\int_{%
\mathcal{Y}}H_{y}\left( V\left( .,y\right) \right) dQ\left( y\right) \right)
.
\end{eqnarray*}

\textbf{Step 3}. From the derivation in Appendix~\ref{app:ContinuousLogit},
we get that%
\begin{eqnarray*}
G_{x}(U\left( x,.\right) ) &=&\frac{\sigma }{2}\log \int_{\mathcal{Y}}\exp 
\frac{U\left( x,y\right) }{\sigma /2}dy\text{ and} \\
H_{y}\left( V\left( .,y\right) \right) &=&\frac{\sigma }{2}\log \int_{%
\mathcal{X}}\exp \frac{U\left( x,y\right) }{\sigma /2}dx
\end{eqnarray*}

Now, in order to get an expression for $\mathcal{I}\left( \pi \right) $ it
remains to compute 
\begin{equation}
\sup_{U\left( x,y\right) }\iint_{\mathcal{X}\times \mathcal{Y}}U\left(
x,y\right) \pi \left( x,y\right) dxdy-\int G_{x}(U\left( x,.\right) )f\left(
x\right) dx  \label{Gstar}
\end{equation}%
and the similar expression on the other side of the market.

By F.O.C.,%
\begin{equation*}
\pi \left( x,y\right) =\frac{f\left( x\right) \exp \frac{U\left( x,y\right) 
}{\sigma /2}}{\int_{\mathcal{Y}}\exp \frac{U\left( x,y\right) }{\sigma /2}dy}
\end{equation*}%
which implies that the value of the problem is infinite unless $\int \pi
\left( x,y\right) dy=f\left( x\right) $, in which case it is%
\begin{equation*}
\left( \sigma /2\right) \iint_{\mathcal{X}\times \mathcal{Y}}\pi \left(
x,y\right) \log \frac{\pi \left( x,y\right) }{f\left( x\right) }dxdy
\end{equation*}%
which is the value of (\ref{Gstar}). A symmetric expression is obtained for
the other side of the market, and finally $\mathcal{I}\left( \pi \right) $
obtains as%
\begin{equation*}
\mathcal{I}\left( \pi \right) =\sigma \iint_{\mathcal{X}\times \mathcal{Y}%
}\log \frac{\pi \left( x,y\right) }{\sqrt{f\left( x\right) g\left( y\right) }%
}\pi \left( x,y\right) dxdy
\end{equation*}%
if $\pi \in \mathcal{M}\left( P,Q\right) $, while $\mathcal{I}\left( \pi
\right) =+\infty $ otherwise.

\textbf{Step 4}. One has 
\begin{eqnarray*}
\mathcal{I}\left( \pi \right) &=&\sigma \iint_{\mathcal{X}\times \mathcal{Y}%
}\log \pi \left( x,y\right) \pi \left( x,y\right) dxdy \\
&&-\left( \sigma /2\right) \int_{\mathcal{X}}\log f\left( x\right) f\left(
x\right) dx-\left( \sigma /2\right) \int_{\mathcal{Y}}\log g\left( x\right)
g\left( x\right) dx
\end{eqnarray*}%
the last two terms do not depend on the particular matching $\pi \in 
\mathcal{M}\left( P,Q\right) $, thus are irrelevant in the expression of the
social welfare, which establishes (\ref{SWF}) and point (i).

(ii) Letting 
\begin{equation*}
a\left( x\right) =\frac{-\sigma }{2}\log \frac{f\left( x\right) }{\int_{%
\mathcal{Y}}\exp U\left( x,y\right) dy}\text{ and }b\left( y\right) =\frac{%
-\sigma }{2}\log \frac{g\left( y\right) }{\int_{\mathcal{X}}\exp V\left(
x,y\right) dx},
\end{equation*}%
one has%
\begin{equation*}
\log \pi \left( x,y\right) =\frac{U\left( x,y\right) -a\left( x\right) }{%
\sigma /2}\text{ and }\log \pi \left( x,y\right) =\frac{V\left( x,y\right)
-b\left( y\right) }{\sigma /2}
\end{equation*}%
and by summation%
\begin{equation*}
\pi \left( x,y\right) =\exp \left( \frac{\Phi \left( x,y\right) -a\left(
x\right) -b\left( y\right) }{\sigma }\right) .
\end{equation*}

(iii) One has%
\begin{equation*}
U\left( x,y\right) =\frac{\sigma \log \pi \left( x,y\right) }{2}+a\left(
x\right) =\frac{\Phi \left( x,y\right) +a\left( x\right) -b\left( y\right) }{%
2}
\end{equation*}%
and similarly%
\begin{equation*}
V\left( x,y\right) =\frac{\Phi \left( x,y\right) -a\left( x\right) +b\left(
y\right) }{2}.
\end{equation*}%
By (\ref{SurplusTransform}), one sees that if man $m$ of type $x$ marries a
woman of type $x$, he gets utility%
\begin{equation*}
u_{m}=\sup_{y^{\prime }\in \mathcal{Y}}\left( U\left( x,y^{\prime }\right)
+\varepsilon _{m}\left( y^{\prime }\right) \right) =U\left( x,y\right)
+\varepsilon _{m}\left( y\right) .
\end{equation*}
\end{proof}

\subsection{Useful lemmas\label{sec:Lemmas}}

We state several useful lemmas which are useful in Sections~\ref%
{sec:Saliency} and \ref{sec:Asympt}, and in the proof of Theorem~\ref%
{thm:Hessian}. First, we need a formula which expresses the affinity matrix
of the rescaled attributes as a function of the affinity matrix between $X$
and $Y$. This is given in the following:

\begin{lemma}
\label{lem:rescale2}For $M$ and $N$ two invertible matrices, one has:%
\begin{equation}
A^{MX,NY}=\left( M^{\prime }\right) ^{-1}A^{XY}N^{-1}.  \label{covariance}
\end{equation}
\end{lemma}

This result should be compared with the expression of the cross-covariance
matrix between $MX$ and $NY$, namely $\Sigma _{MX,NY}=M\Sigma _{XY}N^{\prime
}$. A quick dimensionality check is coherent, as the unit of $A^{XY}$ is the
inverse of the product of the units of $X$ and $Y$, while the unit of $%
\Sigma _{XY}$ is the product of the units of $X$ and $Y$.

\begin{proof}[Proof of Lemma~\protect\ref{lem:rescale2}]
Recall that every affinity matrix $A^{XY}$ is characterized by the fact that:

\begin{equation}
\frac{\partial \mathcal{W}^{P,Q}}{\partial A_{ij}}\left( A^{XY}\right)
=\Sigma _{XY}^{ij}.  \label{firsteq}
\end{equation}

Let $P_{M}$ (resp. $Q_{N}$) be the distribution of $MX$ (resp $NY$). We
therefore have%
\begin{equation}
\frac{\partial \mathcal{W}^{P_{M},Q_{N}}}{\partial A_{ij}}\left(
A^{MX,NY}\right) =\Sigma _{MX,NY}^{ij}=M\Sigma _{XY}^{ij}N^{\prime }=M\frac{%
\partial \mathcal{W}^{P,Q}}{\partial A_{ij}}\left( A^{XY}\right) N^{\prime },
\label{secondeq}
\end{equation}%
where the second equality follows by definition and the third by using (\ref%
{firsteq}). A simple calculation shows that%
\begin{equation*}
\mathcal{W}^{P_{M},Q_{N}}\left( A^{MX,NY}\right) =\mathcal{W}^{P,Q}\left(
M^{\prime }A^{MX,NY}N\right) .
\end{equation*}

Taking the derivative with respect to $A$, yields%
\begin{equation}
\frac{\partial \mathcal{W}^{P_{M},Q_{N}}}{\partial A}\left( A^{MX,NY}\right)
=M\frac{\partial \mathcal{W}^{P,Q}}{\partial A}\left( M^{\prime
}A^{MX,NY}N\right) N^{\prime }.  \label{thirdeq}
\end{equation}

And, by comparing (\ref{secondeq}) and (\ref{thirdeq}), one gets%
\begin{equation*}
\frac{\partial \mathcal{W}^{P,Q}}{\partial A}\left( M^{\prime
}A^{MX,NY}N\right) =\frac{\partial \mathcal{W}^{P,Q}}{\partial A}\left(
A^{XY}\right) .
\end{equation*}

From the strict convexity of $\mathcal{W}^{P,Q}$, we therefore have $%
M^{\prime }A^{MX,NY}N=A^{XY}$, and given that $M$ and $N$ are invertible, it
follows that%
\begin{equation*}
A^{MX,NY}=\left( M^{\prime }\right) ^{-1}A^{XY}N^{-1}.
\end{equation*}%
QED.
\end{proof}

As a consequence of Lemma~\ref{lem:rescale2}, we are able to state that the
results of Saliency Analysis are invariant with respect to a (linear) change
in the measurement units.

\begin{lemma}
\label{lem:invariance}For $\zeta _{i}$ and $\xi _{j}$ two vectors of
positive scalars, let 
\begin{equation*}
\hat{X}_{i}=\zeta _{i}X_{i}\text{ and }\hat{Y}_{j}=\xi _{j}Y_{j}
\end{equation*}%
be the values of partners' attributes measured under different measurement
units. Then the outcome of Saliency Analysis under the new measurement units
coincides with the outcome under the former.
\end{lemma}

\begin{proof}
Saliency Analysis consists in determining the Singular Value Decomposition
of $\Theta =\sigma _{X}A^{X,Y}\sigma _{Y}$ under the old units, and of $\hat{%
\Theta}=\sigma _{\hat{X}}A^{\hat{X},\hat{Y}}\sigma _{\hat{Y}}$ under the new
units. Letting $D_{\zeta }=diag\left( \zeta _{i}\right) $ and $D_{\xi
}=diag(\xi _{j})$, one has%
\begin{equation*}
A^{\hat{X},\hat{Y}}=D_{\zeta }^{-1}A^{X,Y}D_{\xi }^{-1}\text{, }\sigma _{%
\hat{X}}=\sigma _{X}D_{\zeta }\text{ and }\sigma _{\hat{Y}}=D_{\xi }\sigma
_{Y}\text{,}
\end{equation*}%
thus $\hat{\Theta}=\Theta $.
\end{proof}

The next lemma shows that $\mathcal{W}_{1}\left( A\right) $ is strictly
convex.

\begin{lemma}
\label{lem:strictConv}The map $A\rightarrow \mathcal{W}_{1}\left( A\right) $
is strictly convex.
\end{lemma}

\begin{proof}
Consider two matrices $A$ and $\tilde{A}$. Let $\pi $ be the matching
associated to $\Phi \left( x,y\right) =x^{\prime }Ay$, and $\tilde{\pi}$ be
the matching associated to $\Phi \left( x,y\right) =x^{\prime }\tilde{A}y$
(uniqueness of $\pi $ and $\tilde{\pi}$ follows from the uniqueness of the
solution to the Schr\"{o}dinger problem, see R\"{u}schendorf and Thomsen
1993, Theorem 3). Then convexity of $\mathcal{W}_{1}$ implies 
\begin{equation}
\mathcal{W}_{1}\left( \tilde{A}\right) \geq \mathcal{W}_{1}\left( A\right)
+\left\langle \nabla \mathcal{W}_{1}\left( A\right) ,\tilde{A}-A\right\rangle
\label{convexity}
\end{equation}%
where, by the Envelope Theorem, $\nabla \mathcal{W}_{1}\left( A\right) =%
\mathbb{E}_{\pi }\left[ XY^{\prime }\right] $. In order to show strict
convexity, we need to show that equality in (\ref{convexity}) implies $%
A=A^{\prime }$. Assume (\ref{convexity}) holds as an equality. One has%
\begin{eqnarray*}
\mathcal{W}_{1}\left( \tilde{A}\right) &=&\mathcal{W}_{1}\left( A\right)
+\left\langle \nabla \mathcal{W}_{1}\left( A\right) ,\tilde{A}-A\right\rangle
\\
&=&\mathbb{E}_{\pi }\left[ X^{\prime }AY\right] -\mathbb{E}_{\pi }\left[ \ln
\pi \left( X,Y\right) \right] +\mathbb{E}_{\pi }\left[ X\tilde{A}Y^{\prime }%
\right] -\mathbb{E}_{\pi }\left[ X^{\prime }AY\right] \\
&=&\mathbb{E}_{\pi }\left[ X\tilde{A}Y^{\prime }\right] -\mathbb{E}_{\pi }%
\left[ \ln \pi \left( X,Y\right) \right]
\end{eqnarray*}%
This implies that $\pi $ is optimal for the matching problem associated to $%
\Phi \left( x,y\right) =x^{\prime }\tilde{A}y$. Again by the uniqueness of
the solution to the Schrodinger problem mentioned above, it follows that $%
\pi =\tilde{\pi}$, and hence that 
\begin{equation*}
A=\partial ^{2}\ln \pi \left( x,y\right) /\partial x\partial y=\partial
^{2}\ln \tilde{\pi}\left( x,y\right) /\partial x\partial y=\tilde{A}.
\end{equation*}
\end{proof}

The following lemma allows us to characterize the conditional expectations
of the gradient of the log-likelihood.

\begin{lemma}
\label{lem:ANOVA}Let $\pi _{A}\in \mathcal{M}\left( P,Q\right) $ be the
equilibrium matching computed for joint utility function $\Phi _{A}$. Then%
\begin{equation}
\mathbb{E}\left[ \frac{\partial \log \pi _{A}}{\partial A_{ij}}|X=x\right] =%
\mathbb{E}\left[ \frac{\partial \log \pi _{A}}{\partial A_{ij}}|Y=y\right]
=0,  \label{expr2}
\end{equation}%
and%
\begin{equation}
\sigma \mathbb{E}_{\pi }\left[ \frac{\partial \log \pi _{A}\left( X,Y\right) 
}{\partial A_{ij}}\frac{\partial \log \pi _{A}\left( X,Y\right) }{\partial
A_{kl}}\right] =\mathbb{E}_{\pi }\left[ \frac{\partial \log \pi _{A}\left(
X,Y\right) }{\partial A_{ij}}x_{k}y_{l}\right] =\mathbb{E}_{\pi }\left[
x_{i}y_{j}\frac{\partial \log \pi _{A}\left( X,Y\right) }{\partial A_{kl}}%
\right] .  \label{expr1}
\end{equation}
\end{lemma}

\begin{proof}[Proof of Lemma \protect\ref{lem:ANOVA}]
It follows from equation (\ref{Schrodinger}) that%
\begin{equation*}
\sigma \frac{\partial \log \pi }{\partial A_{ij}}\left( x,y\right)
=x_{i}y_{j}-\frac{\partial a}{\partial A_{ij}}\left( x\right) -\frac{%
\partial b}{\partial A_{ij}}\left( y\right) .
\end{equation*}%
But by differentiation of $\int_{\mathcal{Y}}\pi \left( x,y\right)
dy=f\left( x\right) $ with respect to $A_{ij}$, one gets%
\begin{equation*}
\int_{\mathcal{Y}}\frac{\partial \log }{\partial A_{ij}}\pi \left(
x,y\right) \pi \left( x,y\right) dy=0
\end{equation*}%
thus 
\begin{equation*}
\mathbb{E}_{\pi }\left[ \frac{\partial \log }{\partial A_{ij}}\pi \left(
X,Y\right) |X=x\right] =0
\end{equation*}%
which proves (\ref{expr2}). (\ref{expr1}) then follows directly.
\end{proof}

The final lemma in this section shows that the Hessian of $\mathcal{W}$
coincides with the Fisher information matrix $\mathbb{F}$.

\begin{lemma}
\label{lem:hessianW}The Hessian of $\mathcal{W}_{1}$ is given by%
\begin{equation*}
\frac{\partial ^{2}\mathcal{W}_{1}}{\partial A_{ij}\partial A_{kl}}=\mathbb{F%
}_{kl}^{ij}
\end{equation*}%
where the expression of $\mathbb{F}$ is given by \ref{exprF}.
\end{lemma}

\begin{proof}[Proof of Lemma~\protect\ref{lem:hessianW}]
By the envelope theorem, 
\begin{equation*}
\frac{\partial \mathcal{W}_{1}}{\partial A_{ij}}=\int x_{i}y_{j}\pi
_{A}\left( x,y\right) dxdy.
\end{equation*}%
Thus, 
\begin{eqnarray*}
\frac{\partial ^{2}\mathcal{W}_{1}}{\partial A_{ij}\partial A_{kl}} &=&\int
x_{i}y_{j}\frac{\partial \log \pi _{A}\left( x,y\right) }{\partial A_{kl}}%
\pi _{A}\left( x,y\right) dxdy \\
&=&\mathbb{F}_{kl}^{ij}.
\end{eqnarray*}%
where the second equality follows from Lemma~\ref{lem:ANOVA}.
\end{proof}

\subsection{Proof of Theorem~\protect\ref{thm:Hessian}\label%
{sec:proofThmHessian}}

The proof of Theorem \ref{thm:Hessian} relies on the auxiliary results
derived in the previous paragraph. In the sequel, we assume $\sigma =1$; by
positive homogeneity, this is without loss of generality.

\begin{proof}[Proof of Theorem \protect\ref{thm:Hessian}]
Let 
\begin{equation*}
\hat{\pi}\left( x,y\right) =\frac{1}{n}\sum_{k=1}^{n}\delta \left(
x-X_{k}\right) \delta \left( y-Y_{k}\right)
\end{equation*}%
be the distribution of the empirical sample under observation, and $\pi _{A}$
is the equilibrium matching computed for matching utility function $\Phi
_{A} $ (we shall drop the subscript $A$ when there is no ambiguity). Recall
that the (population) affinity matrix $A$ and its sample estimator $\hat{A}$
are respectively characterized by%
\begin{equation*}
\frac{\partial \mathcal{W}_{1}\left( A\right) }{\partial A_{ij}}=\Sigma
_{XY}^{ij}\text{ and }\frac{\partial \mathcal{W}_{1}\left( \hat{A}\right) }{%
\partial A_{ij}}=\hat{\Sigma}_{XY}^{ij}.
\end{equation*}

By the Delta method, we get%
\begin{equation}
\left( \mathbb{F\cdot }\delta A\right) ^{ij}=\int \frac{\partial \log \pi
_{A}}{\partial A_{ij}}\left( \widehat{\pi }-\pi \right) dxdy+o_{D}\left(
n^{-1/2}\right)  \label{deltaMeth}
\end{equation}%
where $\mathbb{F}$ is the Hessian of $\mathcal{W}_{1}$ at $A$, whose
expression is%
\begin{equation*}
\mathbb{F}_{kl}^{ij}=\mathbb{E}_{\pi }\left[ \frac{\partial \log \pi
_{A}\left( X,Y\right) }{\partial A_{ij}}\frac{\partial \log \pi _{A}\left(
X,Y\right) }{\partial A_{kl}}\right]
\end{equation*}%
where $\pi \in \mathcal{M}\left( P,Q\right) $ is the equilibrium matching
computed for the joint utility function $\Phi _{A}$. Further,%
\begin{eqnarray*}
\left( \delta S_{X}\right) ^{ij} &=&1_{\left\{ i=j\right\} }\int
x_{i}x_{j}\left( \widehat{\pi }-\pi \right) dxdy+o_{D}\left( n^{-1/2}\right)
\\
\left( \delta S_{Y}\right) ^{kl} &=&1_{\left\{ k=l\right\} }\int
y_{i}y_{j}d\pi \left( \widehat{\pi }-\pi \right) dxdy+o_{D}\left(
n^{-1/2}\right)
\end{eqnarray*}%
hence%
\begin{equation*}
\mathbb{E}\left[ \left( \mathbb{F\cdot }\delta A\right) _{ij}\left( \delta
S_{X}\right) _{kl}\right] =cov\left( \frac{\partial \log \pi }{\partial
A_{ij}},X_{k}X_{l}\right) 1_{\left\{ k=l\right\} }=0,
\end{equation*}%
where we have used (\ref{expr2}), and similarly, $\mathbb{E}\left[ \left(
\delta A\right) _{ij}\left( \delta S_{Y}\right) _{kl}\right] =0$. This
proves the asymptotic independence between $\delta A$ and $\left( \delta
S_{X},\delta S_{Y}\right) $. The conclusion follows by noting that the
asymptotic variance-covariance matrix of $\delta A$ is $\mathbb{F}^{-1}$,
and that of $\left( \delta S_{X},\delta S_{Y}\right) $ is 
\begin{equation*}
\begin{pmatrix}
\mathbb{K}_{XX} & \mathbb{K}_{XY} \\ 
\mathbb{K}_{XY}^{\prime } & \mathbb{K}_{YY}%
\end{pmatrix}%
.
\end{equation*}
\end{proof}

\subsection{Proof of Theorem~\protect\ref{thm:rankTest}\label%
{sec:proofThmRankTest}}

In order to give asymptotic distributions of matrix estimators, it is
convenient to represent matrices as vectors, an operation which is called 
\emph{vectorization} in matrix algebra. Linear operators acting on these
vectorized matrices will therefore be called \emph{doubly-indexed matrices},
for which we shall use the bold notation to distinguish them from
simply-indexed matrices. If $\mathbb{R}$ is a doubly-indexed matrix, its
general term will be denoted $\mathbb{R}_{kl}^{ij}$, where $ij$ indexes the
lines and $kl$ indexes the columns of $\mathbb{R}$. Then $\mathbb{R}\cdot M$
will denote the (simple) matrix $N$ such that $N^{ij}=\sum_{kl}\mathbb{R}%
_{kl}^{ij}M^{kl}$. We recall the definition of the Kronecker product: for
two matrices $A$ and $B$, $A\otimes B$ is the doubly-indexed matrix $\mathbb{%
R}$ such that 
\begin{equation*}
\mathbb{R}_{kl}^{ij}=A_{ik}B_{jl}.
\end{equation*}

\begin{lemma}
\label{lem:RescaledAffinity}The following convergence holds in distribution
for $n\rightarrow +\infty $:%
\begin{equation*}
n^{1/2}\left( \hat{\Theta}-\Theta \right) \Longrightarrow \mathcal{N}\left(
0,\mathbb{V}\right)
\end{equation*}%
where 
\begin{equation*}
\mathbb{V}=\mathbb{T}_{XY}\mathbb{F}^{-1}\mathbb{T}_{XY}^{\prime }+\mathbb{T}%
_{X}\mathbb{K}_{XX}\mathbb{T}_{X}^{\prime }+\mathbb{T}_{Y}\mathbb{K}_{YY}%
\mathbb{T}_{Y}^{\prime }+\mathbb{T}_{X}\mathbb{K}_{XY}\mathbb{T}_{Y}^{\prime
}+\mathbb{T}_{Y}\mathbb{K}_{XY}^{\prime }\mathbb{T}_{X}.
\end{equation*}
\end{lemma}

\begin{proof}[Proof of Lemma \protect\ref{lem:RescaledAffinity}]
As%
\begin{equation*}
\delta S_{X}^{1/2}=\left( I\otimes S_{X}^{1/2}+S_{X}^{1/2}\otimes I\right)
^{-1}\delta S_{X},
\end{equation*}%
one has%
\begin{eqnarray*}
\delta \Theta &=&\left( S_{Y}^{1/2}\otimes S_{X}^{1/2}\right) \delta
A+\left( S_{Y}^{1/2}A^{\prime }\otimes I\right) \delta S_{X}^{1/2}+\left(
I\otimes S_{X}^{1/2}A\right) \delta S_{Y}^{1/2} \\
&=&\mathbb{T}_{XY}\delta A+\mathbb{T}_{X}\delta S_{X}+\mathbb{T}_{Y}\delta
S_{Y},
\end{eqnarray*}%
where%
\begin{eqnarray}
\mathbb{T}_{X} &=&\left( S_{Y}^{1/2}A^{\prime }\otimes I\right) \left(
S_{X}^{1/2}\otimes I+I\otimes S_{X}^{1/2}\right) ^{-1} \\
\mathbb{T}_{XY} &=&S_{Y}^{1/2}\otimes S_{X}^{1/2} \\
\mathbb{T}_{Y} &=&\left( I\otimes S_{X}^{1/2}A\right) \left(
S_{Y}^{1/2}\otimes I+I\otimes S_{Y}^{1/2}\right) ^{-1},
\end{eqnarray}
\end{proof}

The proof of Theorem~\ref{thm:rankTest} follows as an easy consequence.

\begin{proof}[Proof of Theorem~\protect\ref{thm:rankTest}]
Let%
\begin{equation}
\Omega _{p}=\left( B_{p\perp }\otimes A_{p\perp }^{\prime }\right) \mathbb{V}%
\left( B_{p\perp }\otimes A_{p\perp }^{\prime }\right) ^{\prime }.
\label{exprOmegap}
\end{equation}

By Kleibergen and Paap, Theorem 1, the convergence%
\begin{equation*}
n^{1/2}\hat{T}_{p}\Longrightarrow \mathcal{N}\left( 0,\Omega _{p}\right)
\end{equation*}%
holds for $n\rightarrow +\infty $, and Theorem~\ref{thm:rankTest} follows.
\end{proof}

\section{Computation\label{app:computation}}

Let $a$ and $b$ be the solutions of equation (\ref{Schrodinger}), and
introduce%
\begin{equation*}
\tilde{a}\left( x\right) =\exp \left( -a\left( x\right) /\sigma \right) 
\text{ and }\tilde{b}\left( y\right) =\exp \left( -b\left( y\right) /\sigma
\right)
\end{equation*}%
so equation (\ref{Schrodinger}) rewrites 
\begin{equation}
\pi \left( x,y\right) =\tilde{a}\left( x\right) \tilde{b}\left( y\right)
K\left( x,y\right)  \label{exprPi}
\end{equation}%
where $K\left( x,y\right) =\exp \left( \Phi \left( x,y\right) /\sigma
\right) $, and the system of equations formed by the constraints on the
marginals rewrites%
\begin{eqnarray}
\tilde{a}\left( x\right) &=&f\left( x\right) \left( \int_{\mathcal{Y}}\tilde{%
b}\left( y\right) K\left( x,y\right) dy\right) ^{-1}  \label{margin1} \\
\tilde{b}\left( y\right) &=&g\left( y\right) \left( \int_{\mathcal{X}}\tilde{%
a}\left( x\right) K\left( x,y\right) dx\right) ^{-1}  \label{margin2}
\end{eqnarray}

Note that by (\ref{margin2}), $\tilde{b}$ can be expressed as a function of $%
\tilde{a}$. Then $\tilde{a}$ rewrites as a fixed point equation $\tilde{a}%
=F\left( \tilde{a}\right) $, where $F$ is given by%
\begin{equation*}
F\left( \tilde{a}\right) \left( x\right) =f\left( x\right) \left( \int_{%
\mathcal{Y}}\left( g\left( y\right) \int_{\mathcal{X}}\tilde{a}\left(
x^{\prime }\right) K\left( x^{\prime },y\right) dx^{\prime }\right)
^{-1}K\left( x,y\right) dy\right) ^{-1}.
\end{equation*}

The Iterative Projection Fitting Procedure (IPFP) consists in starting with
some proper choice of $\tilde{a}_{0}\left( x\right) $ that ensures
integrability of $x\rightarrow \tilde{a}\left( x\right) K\left( x,y\right) $%
, and iteratively applying $\tilde{a}_{k+1}=F\left( \tilde{a}_{k}\right) $.
Details and proof of convergence are provided in R\"{u}schendorf (1995);
convergence is very quick in practice.

\newpage

\section{Incorporating singles\label{app:Singles}}

Throughout this appendix, the symbol $\emptyset $ stands for singlehood;
this enlarges the sets of marital choices of men and women, which we denote $%
\mathcal{X}_{0}=\mathcal{X}\cup \left\{ \emptyset \right\} $ and $\mathcal{Y}%
_{0}=\mathcal{Y}\cup \left\{ \emptyset \right\} $. Let $\bar{f}\left(
x\right) $ be the density of mass of men of type $x$, $f_{0}\left( x\right) $
be the density of mass of single men of type $x$, and, as in the rest of the
paper, $f\left( x\right) $ is the density of mass of matched men of type $x$%
, so that $\bar{f}\left( x\right) =f_{0}\left( x\right) +f\left( x\right) $.
Introduce similar notations on the other side of the market: $\bar{g}\left(
y\right) =g_{0}\left( y\right) +g\left( y\right) $, and note that the total
mass of men and women no longer needs to coincide, i.e. in general one has 
\begin{equation*}
\int_{\mathcal{X}}\bar{f}\left( x\right) dx\neq \int_{\mathcal{Y}}\bar{g}%
\left( y\right) dy.
\end{equation*}

The set of acquaintance of man $m$ is now expanded to include singlehood: $%
\{\left( y_{k}^{m},\varepsilon _{k}^{m}\right) ,k\in \mathbb{N}\}$ are now
the points of a Poisson process on $\mathcal{Y}_{0}\times \mathbb{R}$ of
intensity $\lambda _{0}\times e^{-\varepsilon }d\varepsilon $, where for $%
B\subseteq \mathcal{Y}_{0}$%
\begin{equation*}
\lambda _{0}\left( S\right) =1\left\{ \emptyset \in B\right\} +\lambda
\left( B\backslash \left\{ \emptyset \right\} \right)
\end{equation*}%
where $\lambda $ is the Lebesgue measure on $\mathcal{Y}$. As in Appendix~%
\ref{app:ContinuousLogit}, the utility of a man $m$ matching with
acquaintance $k$ is determined at equilibrium by $U\left( x,y_{k}^{m}\right)
+\frac{\sigma }{2}\varepsilon _{k}^{m}$, but $y_{k}^{m}$ can now take value $%
\emptyset $, in which case $U\left( x,\emptyset \right) =\Phi \left(
x,\emptyset \right) $. The indirect utility of man $m$ is thus given by $%
Z=\max_{k}\{U\left( x,y_{k}^{m}\right) +\frac{\sigma }{2}\varepsilon
_{k}^{m}\}$, and one has%
\begin{eqnarray*}
\log \Pr \left( Z\leq c\right) &=&-\iint_{\mathcal{Y}_{0}\times \mathbb{R}%
}1\left( U\left( x,y\right) +\frac{\sigma }{2}\varepsilon >c\right) d\lambda
_{0}\left( y\right) e^{-\varepsilon }d\varepsilon \\
&=&-\exp \left( -c+\log \left( \exp \frac{\Phi \left( x,\emptyset \right) }{%
\sigma /2}+\int_{\mathcal{Y}}\exp \frac{U\left( x,y\right) }{\sigma /2}%
dy\right) \right) ,
\end{eqnarray*}%
so that%
\begin{equation*}
\frac{f_{0}\left( x\right) }{\bar{f}\left( x\right) }=\frac{\exp \frac{\Phi
\left( x,\emptyset \right) }{\sigma /2}}{\exp \frac{\Phi \left( x,\emptyset
\right) }{\sigma /2}+\int_{\mathcal{Y}}\exp \frac{U\left( x,y\right) }{%
\sigma /2}dy}\text{ and }\frac{g_{0}\left( y\right) }{\bar{g}\left( y\right) 
}=\frac{\exp \frac{\Phi \left( \emptyset ,y\right) }{\sigma /2}}{\exp \frac{%
\Phi \left( \emptyset ,y\right) }{\sigma /2}+\int_{\mathcal{X}}\exp \frac{%
V\left( x,y\right) }{\sigma /2}dx}
\end{equation*}%
while%
\begin{equation*}
\pi \left( y|x\right) =\frac{\exp \frac{U\left( x,y\right) }{\sigma /2}}{%
\int_{\mathcal{Y}}\exp \frac{U\left( x,y^{\prime }\right) }{\sigma /2}%
dy^{\prime }}\text{ and }\pi \left( x|y\right) =\frac{\exp \frac{V\left(
x,y\right) }{\sigma /2}}{\int_{\mathcal{X}}\exp \frac{V\left( x^{\prime
},y\right) }{\sigma /2}dx^{\prime }}
\end{equation*}%
hence we see that the observation of $\pi $ identifies $U\left( x,y\right) $
up to an additive term $c\left( x\right) $, and $V\left( x,y\right) $ up to
an additive term $d\left( y\right) $, hence $U$ and $V$ are identified by%
\begin{eqnarray*}
U\left( x,y\right) &=&\sigma /2\left( \log \pi \left( y|x\right) +c\left(
x\right) \right) ,~V\left( x,y\right) =\sigma /2\left( \log \pi \left(
x|y\right) +d\left( y\right) \right) \\
\text{and }\Phi \left( x,y\right) &=&\frac{\sigma }{2}\left( \log \pi \left(
y|x\right) +\log \pi \left( x|y\right) +c\left( x\right) +d\left( y\right)
\right)
\end{eqnarray*}%
where $c\left( x\right) $ and $d\left( y\right) $ are undetermined. This is
precisely the identification achieved in Section~\ref{par:heterog}. The
crucial conclusion is that the observation of singles does not change
anything in the identification of $U$ and $V$. This is a consequence of the
independence of irrelevant alternatives (IIA)\ of the logit model: indeed,
the incentive for remaining single does not affect the odd ratios of the
choices of the partners types. As a result, the distributions of matched men
and women $f\left( x\right) $ and $g\left( y\right) $ may be treated as
exogenous.

Once $U$ and $V$ have been identified, one has%
\begin{equation*}
\frac{f_{0}\left( x\right) }{\bar{f}\left( x\right) }=\frac{\exp \frac{\Phi
\left( x,\emptyset \right) }{\sigma /2}}{\exp \frac{\Phi \left( x,\emptyset
\right) }{\sigma /2}+\exp c\left( x\right) }\text{ and }\frac{g_{0}\left(
y\right) }{\bar{g}\left( y\right) }=\frac{\exp \frac{\Phi \left( \emptyset
,y\right) }{\sigma /2}}{\exp \frac{\Phi \left( \emptyset ,y\right) }{\sigma
/2}+\exp d\left( y\right) }
\end{equation*}%
hence by inversion 
\begin{equation*}
\Phi \left( x,\emptyset \right) =\frac{\sigma }{2}\left( \log \frac{%
f_{0}\left( x\right) }{\bar{f}\left( x\right) -f_{0}\left( x\right) }%
+c\left( x\right) \right) \text{ and }\Phi \left( \emptyset ,y\right) =\frac{%
\sigma }{2}\left( \log \frac{g_{0}}{\bar{g}\left( y\right) -g_{0}\left(
y\right) }+d\left( y\right) \right)
\end{equation*}%
which implies that the observation of single individuals allows one to
identify the reservation utilities. As a result, the \emph{utility surplus
from matching} $\Phi \left( x,y\right) -\Phi \left( x,\emptyset \right)
-\Phi \left( \emptyset ,y\right) $ is identified in the data by%
\begin{equation}
\log \left( \frac{\pi \left( y|x\right) \left( \bar{f}\left( x\right)
-f_{0}\left( x\right) \right) }{f_{0}\left( x\right) }\frac{\pi \left(
x|y\right) \left( \bar{g}\left( y\right) -g_{0}\left( y\right) \right) }{%
g_{0}\left( y\right) }\right)  \label{CSSingles}
\end{equation}%
and \label{page:exAnteUtilities}the ex-ante expected utility surpluses of
men of type $x$ and women of type $y$ are given just as in Choo and Siow by%
\begin{equation}
u\left( x\right) =\log \frac{\bar{f}\left( x\right) }{f_{0}\left( x\right) }%
\text{ and }v\left( y\right) =\log \frac{\bar{g}\left( y\right) }{%
g_{0}\left( y\right) }.  \label{ExAnteUt}
\end{equation}

These formulae are the continuous extensions of the formulae given in Choo
and Siow (2006), where the surplus from matching is identified by $\log
\left( \mu _{xy}^{2}/(\mu _{x0}\mu _{0y})\right) $, where $\mu _{x0}$ and $%
\mu _{0y}$ are respectively the number of single men and women of type $x$
and $y$ respectively, and $\mu _{xy}$ is the number of $xy$ pairs.

\newpage

\section{Further details on the dataset\label{app:Dataset}}

\subsection{Questionnaire about personality and attitudes\protect\footnote{%
The following website:
http://www.centerdata.nl/en/TopMenu/Databank/DHS\_data/Codeboeken/ provides
a link to the complete description of the questionnaire.}\label%
{app:Questionaire}}

\textit{Personality traits, the 16PA scale.}

Now we would like to know how you would describe your personality. Below we
have mentioned a number of personal qualities in pairs. The qualities are
not always opposites. Please indicate for each pair of qualities which
number would best describe your personality. If you think your personality
is equally well characterized by the quality on the left as it is by the
quality on the right, please choose number 4. If you really don't know, type
0 (zero). Scale: 1 2 3 4 5 6 7

TEG1: oriented towards things oriented towards people.

TEG2 slow thinker quick thinker.

TEG3: easily get worried not easily get worried.

TEG4: flexible, ready to adapt myself stubborn, persistent.

TEG5: quiet, calm vivid, vivacious.

TEG6: carefree meticulous.

TEG7: shy dominant.

TEG8: not easily hurt/offended sensitive, easily hurt/offended.

TEG9: trusting, credulous suspicious.

TEG10: oriented towards reality dreamer.

TEG11: direct, straightforward diplomatic, tactful.

TEG12: happy with myself doubts about myself.

TEG13: creature of habit open to changes.

TEG14: need to be supported independent, self-reliant.

TEG15: little self-control disciplined.

TEG16: well-balanced, stable irritable, quick-tempered.

\textit{Attitude towards risk.}

The following statements concern saving and taking risks. Please indicate
for each statement to what extent you agree or disagree, on the basis of
your personal opinion or experience.

totally disagree\ \ \ \ 1 2 3 4 5 6 7\ \ \ \ totally agree

SPAAR1: I think it is more important to have safe investments and guaranteed
returns, than to take a risk to have a chance to get the highest possible
returns.

SPAAR2: I would never consider investments in shares because I find this too
risky.

SPAAR3: if I think an investment will be profitable, I am prepared to borrow
money to make this investment.

SPAAR4: I want to be certain that my investments are safe.

SPAAR5: I get more and more convinced that I should take greater financial
risks to improve my financial position.

SPAAR6: I am prepared to take the risk to lose money, when there is also a
chance to gain money.

\subsection{Construction of the \textquotedblleft Big
Five\textquotedblright\ personality factors\label{app:Variables}}

The DHS panel contains three lists of items that would allow one to assess a
respondent's personality traits.

\begin{enumerate}
\item The first list contains 150 items and refers to the Five-Factor
Personality Inventory measure, developed by Hendriks et al. (1999). This
list was included in a supplement to the 1996 wave.

\item The second list refers to the 16 Personality Adjective (16PA) scale
developed by Brandst\"{a}tter (1988) and was included in the module
\textquotedblleft Economic and Psychological Concepts\textquotedblright\
from 1993 until 2002.

\item From 2003 on, the panel replaced the 16PA scale by the International
Personality Item Pool (IPIP) developed by Golberger (1999). The 10-item list
version of the IPIP scale is used except for the 2005 wave where the 50-item
list was implemented.
\end{enumerate}

Of the three scales, the 16PA scale covers the largest sample of
individuals. For that reason, the 16PA scale was chosen to measure
personality traits. This scale offers the respondents the opportunity to
locate themselves on 16 personality dimensions. Each dimension is
represented by two bipolar scales so that the full scale contains 32 items.
Nyhus and Webley (2001) show that this scale distinguishes 5 factors%
\footnote{%
Using the 1996 wave that contains both the FFPI module and the 16PA module,
Nyhus and Webley (2001) checked the correlation between the 5 factors
identified by the 16PA scale and the (big) five factors identified by the
FFPI. The correlation is generally high though not perfect. This suggests
that both sets of factors assess slightly different aspects of the latent
factors. We followed Nyhus and Webley and use a slightly less general
wording for the various dimensions identified from the 16PA scale.}. They
labeled these factors as: Emotional stability, Extraversion,
Conscientiousness, Agreeableness, and Autonomy. Of the 32 items associated
with the 16PA measure, the first half was asked in 1993, 1995 and each year
between 1997 and 2002 while the other half was asked in 1994 and 1996 only.
Constructing the full scale, therefore, requires losing all respondents but
those who responded in two successive years between 1993 and 1996. To avoid
throwing out too many observations, we constructed the five dimensions using
only those 16 items included in waves 1993, 1995 and 1997-2002. Since
answers given to the same item by the same person in different waves are
strongly correlated (see Nyhus and Webley, 2001), we simply collapse the
data by individual using the person's median answer to each item.

We have constructed our five factors by adding the (standardized) items
identified by Nyhus and Webley (2001) for the respective scales. In other
words, \textquotedblleft Emotional stability\textquotedblright\ is
constructed using items:

\begin{itemize}
\item \textquotedblleft oriented toward reality\textquotedblright
/\textquotedblleft dreamer\textquotedblright ,

\item \textquotedblleft happy with myself\textquotedblright
/\textquotedblleft doubtful\textquotedblright ,

\item \textquotedblleft need to be supported\textquotedblright
/\textquotedblleft independent\textquotedblright ,

\item \textquotedblleft well-balanced\textquotedblright /\textquotedblleft
quick-tempered\textquotedblright ,

\item \textquotedblleft slow-thinker\textquotedblright /\textquotedblleft
quick-thinker\textquotedblright\ and,

\item \textquotedblleft easily worried\textquotedblright /\textquotedblleft
not easily worried\textquotedblright .
\end{itemize}

\textquotedblleft Agreeableness\textquotedblright\ is constructed using
items:

\begin{itemize}
\item \textquotedblleft creature of habit\textquotedblright
/\textquotedblleft open to changes\textquotedblright ,

\item \textquotedblleft slow thinker\textquotedblright /\textquotedblleft
quick thinker\textquotedblright ,

\item \textquotedblleft quiet, calm\textquotedblright /\textquotedblleft
vivid, vivacious\textquotedblright .
\end{itemize}

\textquotedblleft Autonomy\textquotedblright\ is constructed based on:

\begin{itemize}
\item \textquotedblleft direct, straightforward\textquotedblright
/\textquotedblleft diplomatic\textquotedblright ,

\item \textquotedblleft quiet, calm\textquotedblright /\textquotedblleft
vivid, vivacious\textquotedblright\ and,

\item \textquotedblleft shy\textquotedblright /\textquotedblleft
dominant\textquotedblright .
\end{itemize}

\textquotedblleft Extraversion\textquotedblright\ is based on:

\begin{itemize}
\item \textquotedblleft oriented towards things\textquotedblright
/\textquotedblleft towards people\textquotedblright ,

\item \textquotedblleft flexible\textquotedblright /\textquotedblleft
stubborn\textquotedblright\ and,

\item \textquotedblleft trusting, credulous\textquotedblright
/\textquotedblleft suspicious\textquotedblright .
\end{itemize}

\textquotedblleft Conscientiousness\textquotedblright\ is constructed using:

\begin{itemize}
\item \textquotedblleft little self-control\textquotedblright
/\textquotedblleft disciplined\textquotedblright ,

\item \textquotedblleft carefree\textquotedblright /\textquotedblleft
meticulous\textquotedblright\ and,

\item \textquotedblleft not easily hurt\textquotedblright /\textquotedblleft
easily hurt, sensitive\textquotedblright .
\end{itemize}

As a robustness check, we constructed the full scale using the 1993, 1994,
1995 and 1996 waves. We followed Nyhus and Webley (2001) and constructed the
five factors using Principal Component Analysis and varimax rotation on the
five main factors. The correlation between each of the factors we
constructed using only 16 items and the corresponding factor using the full
scale varies between 0.42 for agreeableness and 0.76 for emotional stability.

\pagebreak

\pagebreak

\section{Tables}

\begin{table}[htbp]
\caption{Number of identified young couples and number of young couples with
complete information for various subset of variables.}
\label{tab0}%
\begin{tabular}{rr}
\hline\hline
& N \\ 
Identified couples & 2,897 \\ 
Couples with complete information on: &  \\ 
Education & 2,883 \\ 
The above + Health, Height and BMI$^{a}$ & 1,595 \\ 
The above + Personality traits (Big 5) & 1,258 \\ 
The above + measure of risk aversion & 1,158 \\ \hline
\end{tabular}%
\newline
{\small \ Notes: (1) We have excluded all individuals taller than 210cm or
shorter than 145cm and all individuals lighter than 40kg, no one is heavier
than 200kg in our data. These exclusions represent less than 1 percent of
the sample of adults in the source data. (2) The selected sample for our
analysis is the one from the last row.}
\par
{\small \ a: Excluding health produces exactly the same number of couples at
this stage.}
\par
{\small \ Source: DNB. Own calculation.}
\end{table}

\begin{table}[htbp]
\caption{Sample of young couples with complete information: summary
statistics by gender.}
\label{tab1}\centering\medskip 
\begin{tabular}{lcccccc}
\hline\hline
&  &  &  &  &  &  \\ 
&  & Husbands &  &  & Wives &  \\ \hline
&  &  &  &  &  &  \\ 
& N & mean & S.E. & N & mean & S.E. \\ \hline
&  &  &  &  &  &  \\ 
Age & 1158 & 35.52 & 6.01 & 1158 & 32.78 & 4.84 \\ 
&  &  &  &  &  &  \\ 
Educational level & 1158 & 2.01 & 0.57 & 1158 & 1.87 & 0.57 \\ 
&  &  &  &  &  &  \\ 
Height & 1158 & 182.33 & 7.20 & 1158 & 169.35 & 6.41 \\ 
&  &  &  &  &  &  \\ 
BMI & 1158 & 24.53 & 2.94 & 1158 & 23.44 & 3.83 \\ 
&  &  &  &  &  &  \\ 
Health & 1158 & 3.21 & 0.66 & 1158 & 3.11 & 0.69 \\ 
&  &  &  &  &  &  \\ 
Conscientiousness & 1158 & -0.25 & 0.64 & 1158 & 0.01 & 0.68 \\ 
&  &  &  &  &  &  \\ 
Extraversion & 1158 & -0.12 & 0.69 & 1158 & 0.16 & 0.60 \\ 
&  &  &  &  &  &  \\ 
Agreeableness & 1158 & -0.06 & 0.65 & 1158 & -0.04 & 0.64 \\ 
&  &  &  &  &  &  \\ 
Emotional stability & 1158 & 0.17 & 0.57 & 1158 & -0.19 & 0.53 \\ 
&  &  &  &  &  &  \\ 
Autonomy & 1158 & 0.00 & 0.67 & 1158 & -0.01 & 0.69 \\ 
&  &  &  &  &  &  \\ 
Risk aversion & 1158 & 0.06 & 0.68 & 1158 & -0.12 & 0.88 \\ 
&  &  &  &  &  &  \\ \hline
\end{tabular}
\newline
{\small \ S.E. means Standard Error.}

\end{table}

\begin{landscape}
\begin{table}[htbp]

\caption{\label{tab3} Estimates of the Affinity matrix: quadratic specification (N = 1158).}\centering\medskip

\begin{tabular}{rrrrrrrrrrr} \hline\hline

Wives & Education  & Height  & BMI & Health & Consc.  & Extra. & Agree. & Emotio. & Auto. & Risk \\

Husbands &  &  &   &   &  &  &  &  &  &  \\ \hline \\

Education&\textbf{0.56}&0.02&-0.08&0.02&-0.04&-0.01&-0.03&-0.04&0.05&-0.02\\
Height&0.01&\textbf{0.18}&0.04&-0.01&-0.04&0.05&0.02&0.02&0.02&0.02\\
BMI&-0.05&0.05&\textbf{0.21}&0.01&0.06&0.00&-0.04&0.04&-0.01&-0.01\\
Health&-0.07&0.00&-0.06&\textbf{0.14}&-0.04&0.05&-0.04&0.04&0.02&0.00\\
Consc.&-0.06&-0.03&0.07&0.00&\textbf{0.14}&0.07&0.04&0.06&-0.02&-0.01\\
Extra.&0.01&-0.02&0.05&0.02&-0.06&0.02&-0.02&-0.01&-0.03&-0.05\\
Agree.&0.00&0.01&-0.08&0.02&\textbf{0.13}&\textbf{-0.14}&0.02&0.11&-0.09&-0.04\\
Emotio.&0.03&0.00&\textbf{0.12}&0.04&\textbf{0.21}&0.05&-0.03&-0.04&0.08&0.01\\
Auto.&0.02&0.00&0.00&0.01&\textbf{-0.11}&\textbf{0.11}&-0.04&0.03&-0.09&0.01\\
Risk&0.00&0.02&-0.03&0.02&0.01&-0.01&-0.01&-0.05&0.05&\textbf{0.11}\\

\hline

\end{tabular} \\
{\small \ Note: Bold coefficients are significant at the 5 percent level.}

\end{table}

\end{landscape}

\begin{landscape}
\begin{table}[htbp]

\caption{\label{tab5a} Share of observed joint utility explained.}\centering\medskip

\begin{tabular}{rrrrrrrrrrr} \hline\hline
& I1  &  I2 & I3  & I4  & I5 & I6 & I7 & I8 & I9 & I10 \\

Share of joint utility explained&27.98***&16.60***&14.20***&10.07***&9.18***&8.51***&6.24***&4.14***&2.09&0.99\\
Standard deviation of shares&2.25&1.55&1.59&1.54&1.68&1.48&2.58&1.91&2.26&1.03\\

\hline

\end{tabular} \\
{\small \ I1-I10 indicates the 10 indices created by the Singular Value Decomposition of the affinity matrix.}
\par
{\small \*** significant at 1 percent.}

\end{table}

\end{landscape}

\begin{landscape}
\begin{table}[htbp]

\caption{\label{tab5} Indices of attractiveness.}\centering\medskip

\begin{tabular}{rrrrrrrrrrr} \hline\hline
& I1  &   & I2 &  & I3   &  \\

Attributes & M & W & M  & W  & M & W  \\ \hline \\

Education&\textbf{0.97}&\textbf{0.96}&0.15&0.21&-0.01&-0.02\\
Height&0.02&0.04&0.02&0.05&-0.39&-0.27\\
BMI&-0.16&-0.19&0.41&0.51&-0.35&\textbf{-0.56}\\
Health&-0.08&0.02&-0.20&0.02&-0.04&-0.02\\
Consc.&-0.17&-0.14&0.37&\textbf{0.82}&0.04&0.39\\
Extra.&0.01&-0.02&-0.08&-0.02&-0.17&\textbf{-0.59}\\
Agree.&-0.02&-0.05&0.16&0.00&\textbf{0.75}&0.17\\
Emotio.&-0.01&-0.09&\textbf{0.71}&0.01&-0.15&0.22\\
Auto.&0.05&0.08&-0.30&0.18&-0.33&-0.17\\
Risk&0.02&-0.02&-0.00&-0.02&-0.06&-0.11\\

\hline

Cum. share  &       27.98 &   &    44.58 &  &     58.78 & \\

\hline

\end{tabular} \\
{\small \ Notes: I1-I3 are the respective indices. M means men and W means women.}
\par
{\small \ Bold coefficients indicate coefficients larger than 0.5.}

\end{table}

\end{landscape}

\end{document}